\documentclass[11pt,preprint,graphicx]{aastex}
\usepackage{graphicx}

\def\etal{\it et al. \rm }

\begin{document} 

\title{The Structure of Galaxies: I. Surface Photometry Techniques}

\author{J. Schombert$^{A,B}$, A. K. Smith$^{A,C}$}
\affil{$^A$Department of Physics, University of Oregon, Eugene, OR USA 97403}
\affil{$^B$jschombe@uoregon.edu}
\affil{$^C$annas@uoregon.edu}

\begin{abstract}

\noindent This project uses the 2MASS all-sky image database to study the
structure of galaxies over a range of luminosities, sizes and morphological
types.  This first paper in this series will outline the techniques,
reliability and data products to our surface photometry program.  Our
program will analyze all acceptable galaxies (meeting our criteria for
isolation from companions and bright stars) from the Revised Shapley-Ames
and Uppsala galaxy catalogs.  Resulting photometry and surface brightness
profiles are released using a transparent scheme of data storage which
includes not only all the processed data but knowledge of the processing
steps and calibrating parameters.

\end{abstract}

\section{Introduction}

The photometry and structure of galaxies are key testing predictions from
galaxy formation scenarios and in understanding the later evolution of
galaxies.  Photometry reveals the amount and distribution of stellar mass
(with proper SED modeling to determine $M/L$), where the stars are the
primary baryonic component for most galaxy types. Multi-color photometry
explores the properties of the underlying stellar population (chemical and
star formation history).  The distribution of light, as given by galaxy
structure, is used to deduce the size and shape of the gravitational
potential, a primary variable to the fundamental plane (Kormendy 1977).
However, where recent semi-analytic formation models have successfully been
applied to galaxy morphology, chemical histories, luminosity function, gas
fractions and galaxy sizes (Cole \etal 2000), the theoretical world has
been strangely quiet on predictions for the structure of galaxies
themselves.

The use of structural information varies with galaxy type.  For example,
structure of early-type galaxies is critical in distinguishing between cold
dissipationless collapse versus hierarchical merging scenarios (Conselice
2008).  These same models predict the ratio of spheroid to disk galaxies,
as well as the distribution of bulge-to-disk ratios (Almeida \etal 2007),
so the structure of early-type galaxies is relevant to these predictions and
their structural parameters are of importance to dark matter studies.

Both photometric and structural analysis are the domain of galaxy surface
photometry (de Vaucouleurs 1948, Fish 1964, Freeman 1970).  Surface
photometry, the study of extended objects in the sky, has the primary goal
of quantifying the 2D light distribution of galaxies (Milvang-Jensen \&
Jorgensen 1999) and ultimately reducing the isophotes to a 1D set of
parameters.  Reduction of galaxy images to surface photometry is aided by
the fact that, for most morphological types, galaxy isophotes are closely
approximated by ellipses.  Elliptical isophotes, of course, simply reflect
the condition that underlying stellar orbits in galaxies are Keplerian and,
even for irregular galaxies, an ellipse is a fair description for isophotes
lacking the distorting effects of ongoing star formation or dust
extinction.

The reduction of galaxy images to surface brightness profiles was, in the
past, a time consuming process due to the amount of user interaction
required by the nature of imaging processing.  However, advances in
computer languages and processing power has simplified many of the initial
stages of image analysis to the point where it is currently possible to
reduce extremely large numbers of galaxies in less time than that needed to
actually construct and test the algorithms.  However, galaxies are not
uniform in their appearance, and forcing the reduction of their light
distributions (either at the 2D or 1D levels) often leads to a loss of
potentially valuable information (this is particularly true of late-type
galaxies).

The goal of this project is the map the structure of galaxies over a large
range of luminosities, sizes and morphological types.  To this end, we have
extracted a sample of large angular-sized ($D > 1$ arcmin) galaxies from
the 2MASS near-IR all-sky survey.  The 2MASS survey is ideal as near-IR
wavelengths are dominated by stellar light, the primary baryonic component
in galaxies, and minimizes distorting effects due to gas and dust
extinction.  While 2MASS images are short in exposure time (effectively 7.8
secs), there is sufficient S/N to achieve faint surface brightness levels
for axial symmetric objects.  In addition, 2MASS images provide
simultaneous coverage in $JHK$ (i.e., multi-color photometry) in order to
study the color gradients, important to population studies.

There is no attempt to obtain a complete sample of galaxies in this
project, and galaxies with companions or nearby bright stars are rejected
due to complications from overlapping isophotes.  Our goal is to analyze as
many galaxies in the Revised-Shapley Ames catalog (RSA, a catalog selected
by luminosity) and Uppsala Galaxy Catalog (UGC, an angular limit catalog)
which satisfy our criteria of isolation from foreground or background
objects.  Thus, this series of papers will explore the surface photometry
of galaxies in the 2MASS image database over a full range of galaxy
morphological types, starting with the early-type systems and proceeding
through the Hubble sequence.  Our first papers will outline our techniques
and statistical methods, focusing in particular on the limits and errors to
2MASS galaxy photometry.  Later papers will address each morphological
class and the discoveries we make in each category.

We also introduce, in this series of papers, a new avenue for published
data access and transparency.  We will present all the data contained in
our study with the full set of calibrated data along with the scripts used
to transform raw image numbers into astronomically meaningful values.  This
allows any researcher access to the parameters for any galaxy we have
reduced and, most importantly, to cross check our results.  In addition, we
provide all the reduced numbers in XML format, which allows a user to ask
for specific parameters seamlessly across the dataset (i.e., make your own
tables).  Full access to the data can be found at
http://abyss.uoregon.edu/ $\sim$js/sfb.

\section{2MASS Imaging Database}

The 2MASS project was a NASA ground-based, all-sky, near-IR sky survey
(Skrutskie \etal 2006).  2MASS uniformly scanned the entire sky using two
1.3-m telescopes (north KPNO and south CTIO).  Each telescope was equipped
with a three-channel camera, where each channel consisting of a 256x256
HgCdTe detector.  Each camera was capable of observing the sky
simultaneously at $J$ (1.25 microns), $H$ (1.65 microns), and $K$ (2.17
microns).  The 2MASS arrays imaged the sky in a drift-scan mode.  Each
final pixel consisted of six pointings on the sky for a total integration
time of 7.8 sec per pixel.  The final image frames have a plate scale of
one arcsec per pixel.

Any region of the sky is available from 2MASS's Atlas Image server.
However, these images are in the form of sky strips, which rarely conform
to the centroids of bright galaxies.  Lacking a tool at the 2MASS website
to merge sky frames, a short image construction script was developed that
introduces a novel combination of network and image processing tools.  The
script takes a galaxy's name from a user supplied list.  It then parses
that name through the NED (NASA's Extragalactic Database) server to extract
correct RA and Dec information.  With these coordinates, the script then
accesses the 2MASS Atlas Image server extracting the four sky strips to the
north, east, west and south of the galaxy center.  These four images are
than stitched together, using the geometric information found in each
frame's header, to produce a 512x512 or 1024x1024 sized raw image (image
size is irrelevant to the software, these numbers were simply a
historically pleasing choice with sufficient sky around each galaxy).
Galaxies greater than 5 arcmins in diameter were excluded from the study
since the night sky would vary between sky strips for objects larger than a
few scans.  These image sizes allowed for a sufficient number of pixels to
determine sky values, and to exclude any galaxies with companions.

\begin{figure}[!ht]
\centering
\includegraphics[scale=0.5,angle=0]{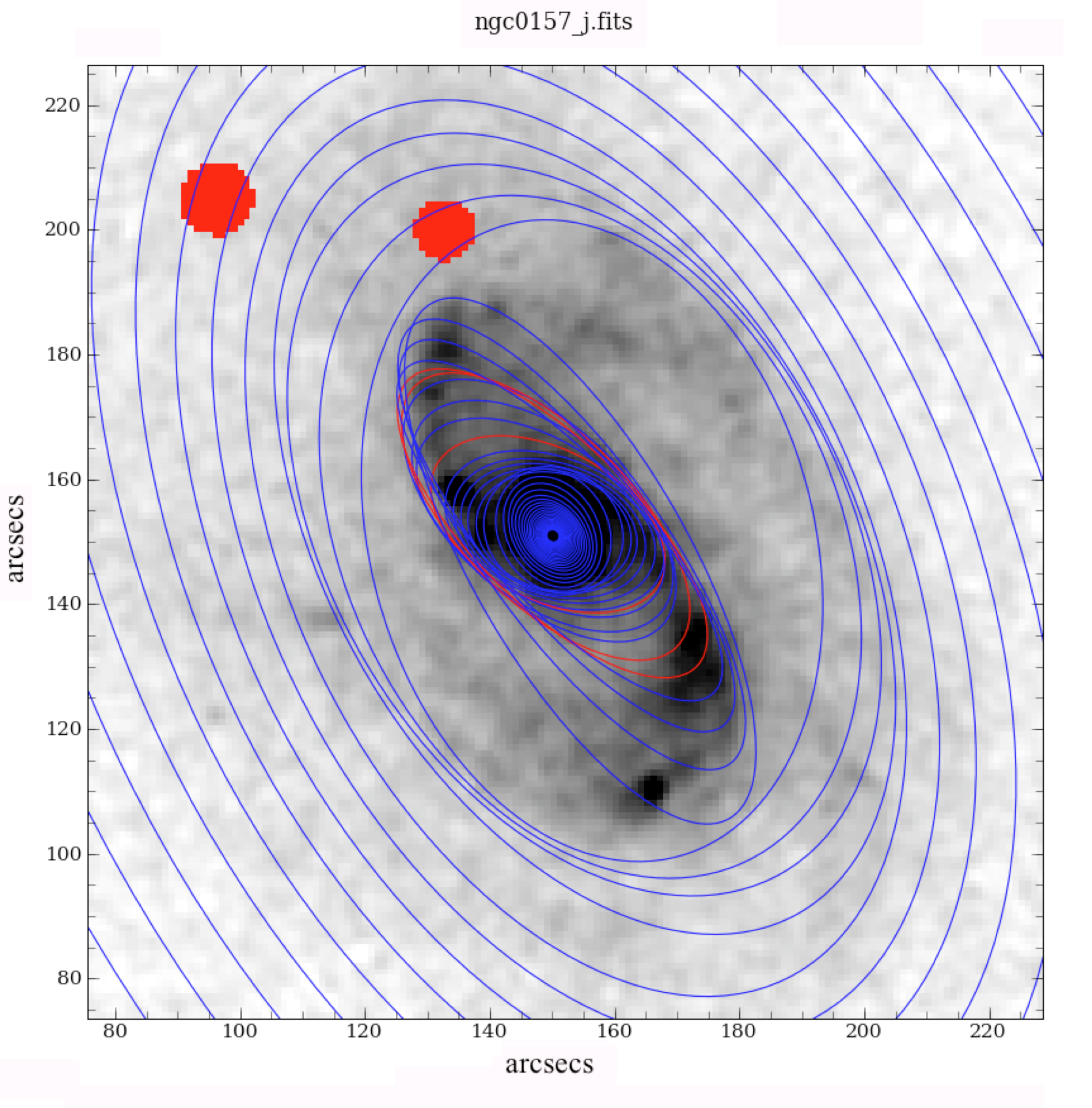}
\caption{\small The 2MASS $J$ image of the inner 150 arcsecs of late-type
spiral NGC 157.  Axes are in arcsecs from the corner of the frame.  Best
fit ellipses are shown in blue.  Ellipses marked red are the best fit at
the 50 iteration limit.  Red areas are stars removed by the processing
pipeline.  Even for a highly irregular disk galaxy, ellipses are a fair
description of the isophotes and closely follow the changing structure of
the galaxies inner regions.  Aperture photometry would recover the missing
luminosity that the isophotes averaged over.
}
\end{figure}

Calibration was provided by the 2MASS project, although we confirmed these
values by comparison with galaxy aperture photometry values in the literature.
The supplied calibrations were never in difference with aperture values by
more than 2\% and do not contribute a notable fraction of the error
budget.  Other relevant information in the scan headers provided
information of the original counts and sky conditions.  This information
was passed to the processing pipeline to be incorporated into our error
analysis.

\section{Data Reduction}

Our surface photometry pipeline used the galaxy photometry package
ARCHANGEL (Schombert 2007).  ARCHANGEL is a long running software project
by one of us (JS) that has its origins in photographic imaging data of
the Second Palomar Sky Survey.  Its history has extended over four computer
languages and five different operating systems.  While numerous tools and
GUI's have been developed to provide increasingly sophisticated image
analysis capability, the core elements of galaxy surface photometry are 1)
global frame cleaning, flattening and sky determination, 2) isophotal
fitting (typically with ellipses) and 3) reduction of the 2D information to
1D surface brightness profiles, aperture magnitudes and structural
parameter (compactness, asymmetry, etc.).  Interpretation of the surface
brightness profiles deserves a separate enterprise that involves the
specifics of fitting functions and kinematic models and will be addressed
in a separate paper.

Image preparation is the first step in an galaxy photometry project.
Initial work can range from correcting for instrumental distortions
(flattening, dark current, cosmic rays, dead pixels, etc.) to
characterizing the properties of a frame (readout noise, geometric
distortion, etc.).  For this project, the 2MASS project provides
calibrated, flattened, kernel smoothed, sky-subtracted images.  There are
artifacts due to instrument irregularities (e.g., latent images), however,
these are treated in the same manner as stellar and other non-galaxian
objects.

\subsection{Sky Determination}

Good galaxy photometry is critically dependent on accurate knowledge of the
sky value.  For, in the outer regions of galaxies, the galaxy luminosity
per pixel is a small percentage of the sky flux.  Thus, errors in the sky
value will dominate Poisson noise or calibration errors.  Errors in the sky
value are limited by 1) quality of the flatness to an image and 2) its
proper assessment using regions that are free of artifacts, foreground
stars and galaxy or background light.

With regard to image quality, the 2MASS observing scheme (co-added drift
scans) produced extremely flat, uniform images.  Our estimates, during the
initial stages of this project, was that the sky values had a mean
variation on small scales (estimated from hundreds of images) of only
0.1\%.  No coarse gradients or other large scale features were detected.

This means that the greatest source of uncertainty in determining a global
sky value is careful selection of the proper pixels free of any
contaminating objects.  For this reason, based on experience working with
low surface brightness galaxies data from optical CCD's, the use of sky
boxes is recommended.  Sky determination by sky boxes involves an
interaction with the data frame where the user selects areas, clear of
obvious stellar or galaxy sources, which are then summed and averaged.
The pixels in each sky box are also clipped at the 3$\sigma$ level to
eliminate 'hot' pixels and cosmic rays.

This procedure has the advantage in that if ten or more sky boxes are
measured, then the user not only determines the mean and standard deviation
for every box but also extracts the mean of the mean values for all the
boxes, and the standard deviation for that total average.  It is the
variation between the mean box values that is the true measure of the
correct sky value and is the primary estimate of error at low galaxy light
levels.

\begin{figure}[!ht]
\centering
\includegraphics[scale=0.3,angle=0]{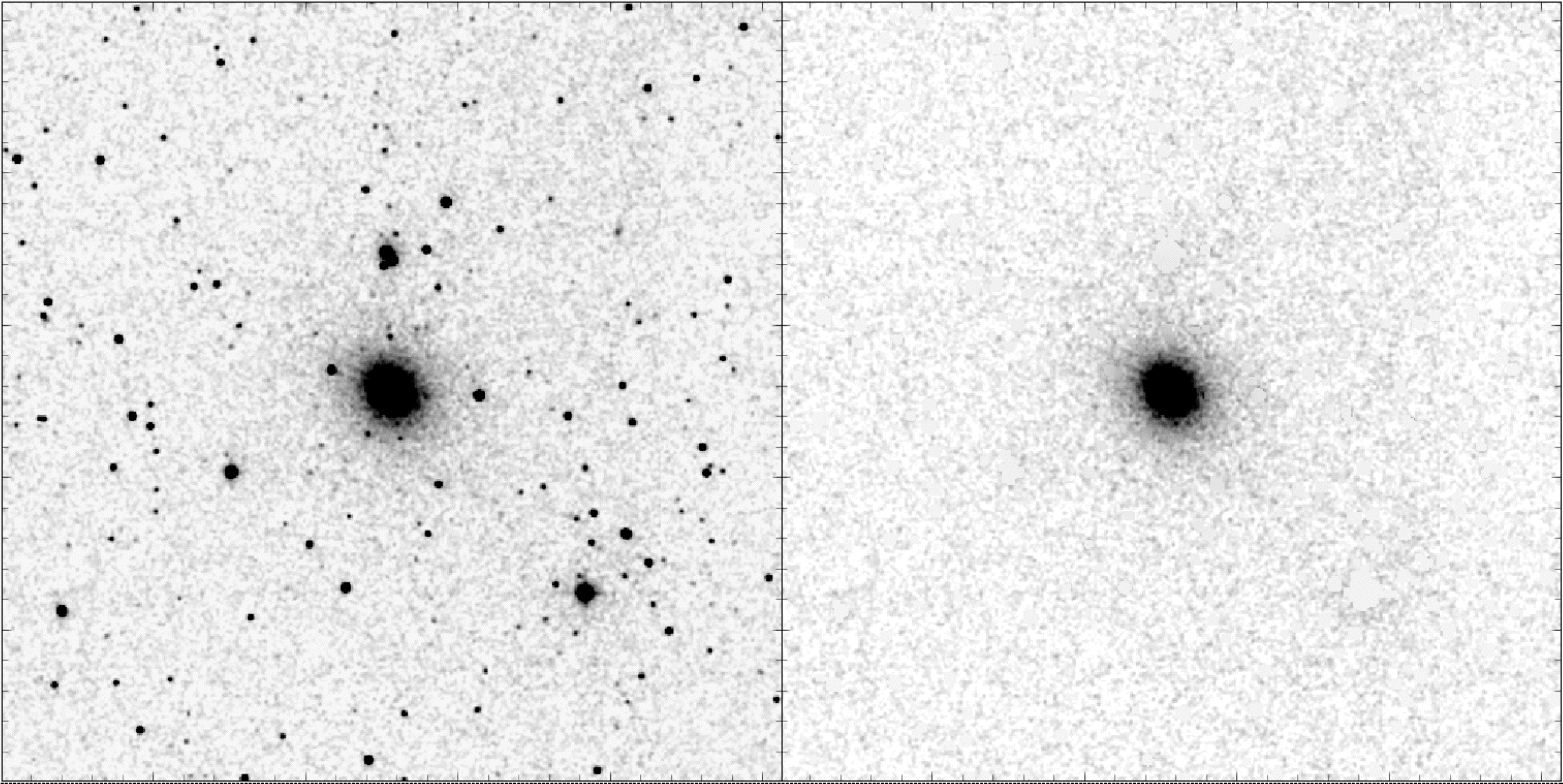}
\caption{\small An example of the cleaning algorithm for early-type galaxy
NGC 3087.  The raw image is on the left.  Each frame is 10 arcmins to a
side.  The algorithm iterates on the
best fit ellipse to subtract all pixels (and a growth radius) that are
5$\sigma$ above the mean.  After the ellipse fitting routine has measured
the entire galaxy, a second routine replaces the removed pixels with the
interpolated intensities from the elliptical isophotes.  The final,
cleaned image is shown on the right.
}
\end{figure}

The 2MASS image frames are sky subtracted, i.e. the sky value is zero. 
The typical standard deviation on the sky was 0.11 DN (which corresponds to
23.2 $J$ mag arcsecs$^{-2}$) and only varied by 0.05 between the frames
examined for this project.  Our own estimates of the sky value using our
sky box method only deviated from zero by 0.04 DN, less than 30\% the
variance on the 2MASS project's sky value.  While we adopted our sky
values, there was no significant difference than the value of zero provided
by 2MASS.  The actual sky brightness was between 15.5 and 16.3 $J$ mag
arcsecs$^{-2}$ for the frames we examined, remarkably consistent
considering the variability of the near-IR sky.

\subsection{Isophotal Analysis}

Once sky has been determined, and gross contaminating features have been
removed, the next step towards surface photometry is the extraction of
isophotal values as a function of radius.  As mentioned above, isophotes in
galaxies are typically elliptical in shape.  This is a convenient
description for isophotes as an ellipse only has its center, position angle
and eccentricity as variables.

\begin{figure}[!ht]
\centering
\includegraphics[scale=0.7,angle=0]{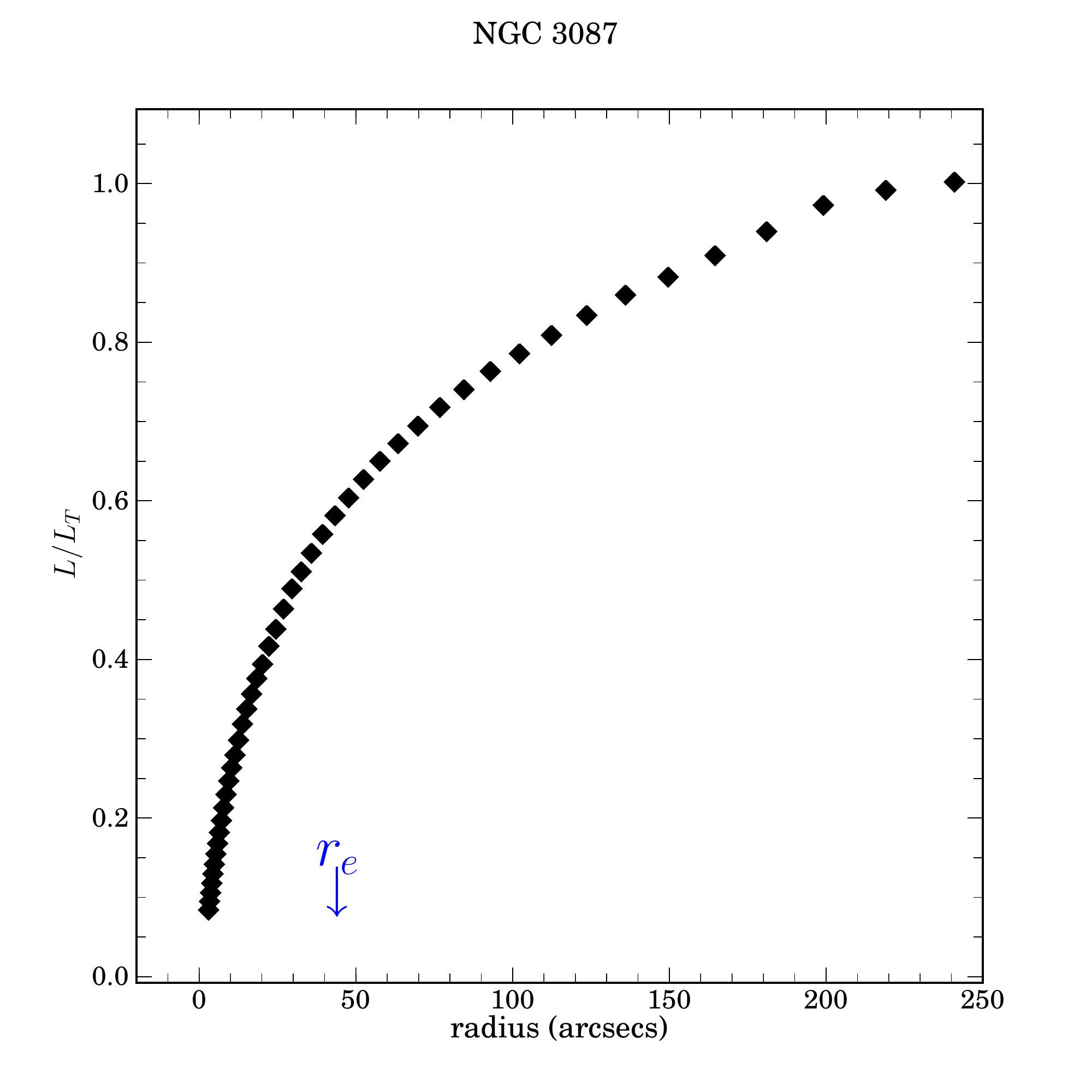}
\caption{\small Elliptical aperture luminosity as a function of radius for
the early-type galaxy NGC 3087.  The galaxy has a total isophotal radius of
approximately 250 arcsecs, but over 50\% the total luminosity of the galaxy
is outside 50 arcsecs.  This is typical of early-type galaxies and
demonstrates that determining the correct total stellar mass of a galaxy is
very sensitive to photometry in the faintest portion of a galaxy's light
distribution, its outer envelope.  The effective radius, based on fits by a S\'{e}rsic,
is marked and is a close match to the one-half luminosity point.
}
\end{figure}

Fitting a best ellipse to a set of intensity values in a 2D image is a
relatively straight forward technique that has been pioneered by Cawson
\etal (1987) and refined by Jedrzejewski (1987) (see also an excellent
review by Milvang-Jensen \& Jorgensen 1999).  The core routine from these
techniques (PROF) was eventually adopted by STSDAS IRAF (i.e. ELLIPSE).
The primary fitting routine for this project follows the same techniques
(in fact, uses much of the identical FORTRAN code from the original GASP
package of Cawson with some notable additions).

These codes start at some intermediate distance from the galaxy core with
an estimated x-y center, position angle and eccentricity then begin
sampling the pixel data around the given ellipse.  The variation in
intensity values around the ellipse can be expressed as a Fourier series
with small second order terms.  Next, an iterative least-squares procedure
adjusts the ellipse parameters searching for a best fit, i.e. minimized
coefficients.  There are several halting conditions, such as maximum number
of iterations or minimal/extreme change in the coefficients, which then
moves the ellipse outward for another round of iterations.  Once a stopping
condition is met (edge of the frame or sufficiently small change in the
isophote intensity), the routine returns to the start radius and completes
the inner portion of the galaxy.  A side benefit to above procedure is that
the cos(4$\theta$) components to each isophote fit are easily extracted,
which provides a direct measure of the geometry of the isophote (i.e. boxy
versus disk-like, Jedrzejewski 1987).

One addition, from the original routines, is the ability to clean (i.e.
mask) pixels along an isophote.  Basically, this routine first allows a few
iterations to determine a mean intensity and RMS around the ellipse.  Any
pixels above (or below) a multiple of the RMS (i.e. 3$\sigma$) are set to
not-a-number (NaN) and ignored by further processing.  Due to the fact that
all objects, stars and galaxies, have faint wings, a growth factor is
applied to the masked regions.  While this process is efficient in
early-type galaxies with well defined isophotes, it may be incorrect in
late-type galaxies with bumpy spiral arms and HII regions.  The fitting
will be smoother, but the resulting isophotometry will be underestimated.

An example of the above procedure is shown in Figure 1, the
$J$ image of the late-type galaxy NGC 157.  Converging ellipses are shown
in blue, best fits (but exceeding a maximum iteration value) are shown in
green.  Even though this galaxy has axial symmetric, there are several
regions of star formation which could distort the ellipse fitting.
However, the subtraction algorithm works through the irregular light
distribution to find reasonable fits (masked pixels are later restored to
the mean isophotal value 
for aperture luminosities determination).  For early-type galaxies
this is not an issue, but we will revisit this problem during our analysis
of late-type systems.

\subsection{Aperture Photometry}

Historically, determining isophotal or aperture magnitudes has required the
use of curves of growth (de Vaucouleurs 1977), and assumptions to the overall
structure of galaxies, in order to capture the luminosity in the outer
regions.  The movement from photographic materials to digital imaging
results in a more accurate measure of the outer luminosity of galaxies and
curves of growth are no longer required to obtain total magnitudes.

This is not to say that galaxy photometry is not without its challenges.
For example, an obvious problem to measuring galaxy magnitudes is
separating galaxy light from foreground stars or background galaxies.
Fortunately, a byproduct of the ellipse fitting routine is the elimination
of non-galaxy sources by the cleaning algorithms.  The ellipse fitting
routine replaces bad pixels with a non-value (NaN).  It is a simple
procedure to replace those masked pixels with intensity values interpolated
from the nearest ellipse values.  This results in a more accurate
measurement of the galaxy light through any aperture.

An example of the output from the cleaning procedures is shown in Figure
2 for the early-type galaxy NGC 3087.  The visual difference is
impressive, although the difference in the total flux measurements with
versus without star subtraction is only 3\% since the stars only obscure a
very small portion of the typical galaxy.  While the difference is small,
there is still merit in using star subtracted frames for a clearer view of
the morphological appearance of galaxies and, for late-type galaxies, a
clearer view of the color distribution.

The greatest challenge in galaxy photometry is correctly determining the
total luminosity since this is equivalent to the total stellar mass for a
galaxy (using an assumed $M/L$ and ignoring internal extinction effects).
The difficulty is that a significant portion of a galaxy's total light is
in the outer regions, where the S/N per pixel is the lowest.  This is
demonstrated in Figure 3, a plot of elliptical aperture
magnitude for the galaxy NGC 3087 as a function of radius (arcsecs).  The
luminosity is plotted as the ratio of the aperture luminosity to the total
luminosity.  One can see that over 50\% of the total luminosity of a galaxy
is contained in the outer 95\% of a galaxy's area.

Since a majority of the luminosity of galaxy is in the region of its light
distribution with the lowest S/N per pixel, this places a high burden on
the algorithms that are attempted to measure this luminosity.  In
particular, the pixel-to-pixel noise will quickly dominate the error
budget.  Traditional methods of plotting luminosity as a function of radius,
and mapping the total magnitude to a convergence point, will frequently
fail when small errors in the sky value produce a divergence for the
aperture luminosities.

The procedure adopted by this project is to allow the surface photometry to
guide the aperture pixels, where the apertures are defined by the best-fit
ellipses (a visual example is found in Figure 9 of Schombert 2007).  In
other words, in the outer regions, where the S/N per pixel is low, rather
than summing the pixels, our algorithms use the mean isophotal value at
that radius for the pixel's intensity.  This procedure assumes strong
symmetry to the galaxy's light distribution and will be more effective for
early-type galaxies than late-type galaxies.

As discussed in our photometry package paper (Schombert 2007), it was rare
to fail to find convergence of the luminosity-radius plot using this
technique.  The final value is calculated by fitting an asymptotic function
to the data points at large radius, with the calculated magnitude stored in
our XML files as {\tt tot\_sfb\_mag}.  An example of this procedure, for
clarity, is found in Appendix B.

Extracted during the same processing for total magnitude is the total size
of the galaxy, defined as the radius where the total luminosity is reached.
However, this value is extremely uncertain as even small errors in the
total magnitude reflect into large changes in the radius where the total
magnitude is reached.  However, total radius is correlated with other
characteristic scalelengths (such as half-light radius, isophotal radius
and effective radius) which will be discussed in a later paper.

\subsection{Surface Photometry}

The last step in characterizing the structure of galaxies is conversion of
the 2D isophotes into a 1D surface brightness profile.  The elliptical
isophotes are converted to luminosity density (magnitudes arcsecs$^{-2}$)
by subtracting sky, dividing by the plate scale (squared) and adding the
appropriate photometric constant.  For this study, all structural
parameters are determined in the uncorrected units (i.e., surface
brightness without corrections for galactic extinction or cosmological
dimming, radii in arcsecs rather than kpc).

As with all reduction techniques, the goal is to summarize complex data
(e.g., an image) into some set of parameters which are representative of
the original data, but which allows for comparison to other data.  It is
also assumed that there is no critical loss of information in the reduction
process or that the loss is not relevant to the science questions being
addressed.  With respect to surface photometry, this tension between the
images and reduced data is reflected in the loss of morphological
information with the conversion of a galaxy's light distribution to a
surface brightness profile.  However, even in this circumstance, some
morphological information is preserved, e.g. a bulge+disk versus a
power-law profile.

A second concern for data reduction techniques is repeatability.  In our
current era of fast digital imaging, repeatability for the original images
is high, varying only as given by the observing conditions (this differs
from the photographic era where the `art' of astrophotography produced a
wide range in image quality).  The many steps between image and surface
brightness profile raises the concern that different methods would produce
wildly different results.

Repeatability usually depends on comparison of the same galaxy from
different projects.  Comparison to 2MASS profiles is not meaningful as it
has already been demonstrated (Schombert 2007) that their reduction pipeline
improperly reduced their galaxy images by distorting the ellipse fitting
and underestimating the sky value (see Appendix A and Schombert 2011).

In order to demonstrate repeatability we have compared our current data
with a set of ellipticals reduced by one us (JS) for a study of cluster
galaxies (Schombert 1986).  One such comparison is shown in Figure
4 where we overlay the $V$ and $J$ surface brightness
profiles for the Coma elliptical NGC 4881 (the $V$ data has been corrected
for $V-J$=2.57, taken from NED).  NGC 4881 is a popular surface brightness
target as its elliptical shape is close to a perfect circle and its profile
is r$^{1/4}$ over a large range of surface brightness.  The close
correspondence between the two profiles is remarkable since the $V$ data is
taken from PDS scans of Palomar Schmidt photographic material over 20 years
ago.  There is a slight difference in slope for the two profiles, however,
due to the blue nature of this slope, this is more than likely the result
of a small metallicity gradient (i.e., negative $V-J$ slope).

\begin{figure}[!ht]
\centering
\includegraphics[scale=0.8,angle=0]{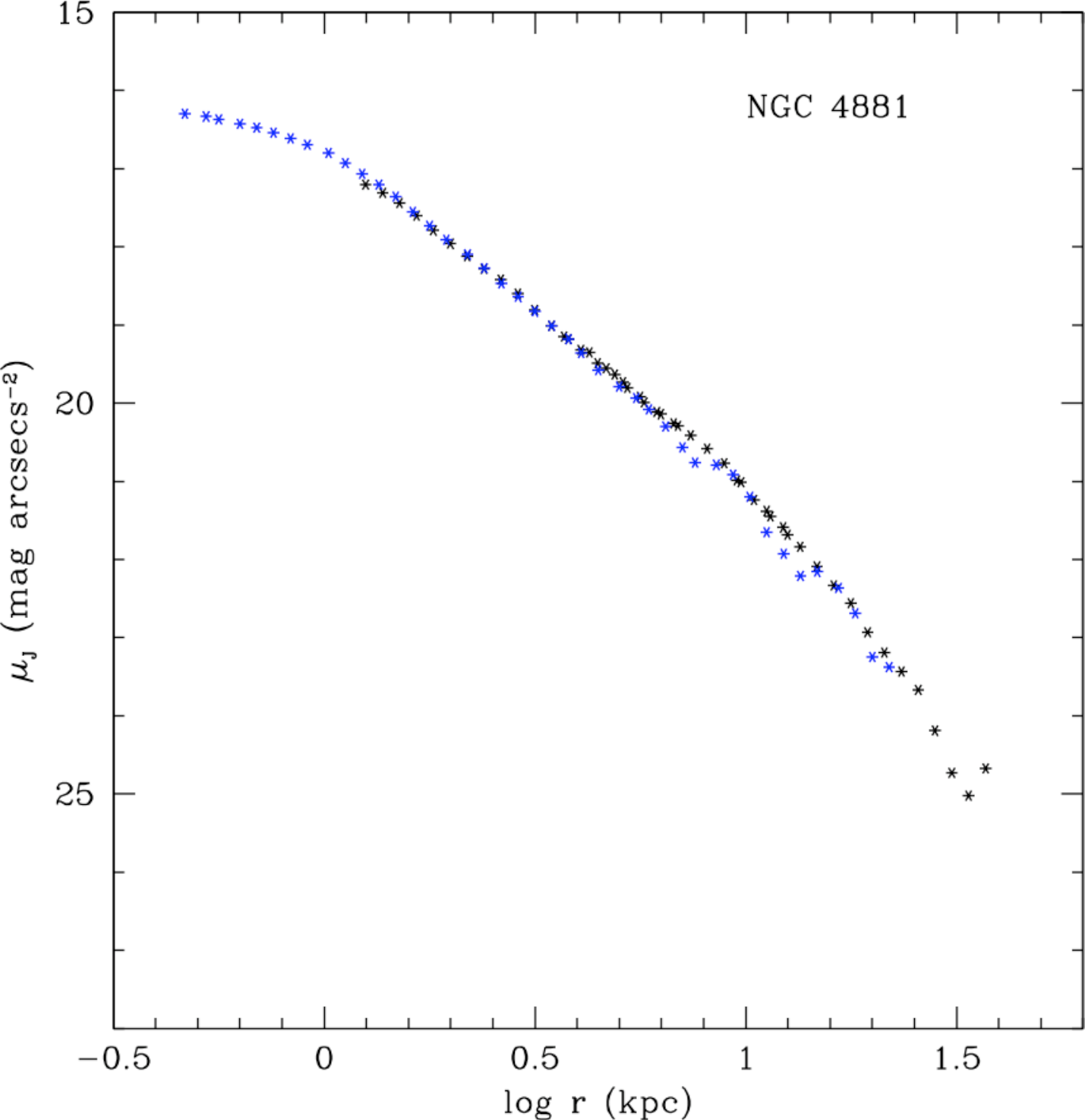}
\caption{\small A comparison of the surface brightness profile for NGC
4881.  The blue data is the 2MASS $J$ data from this project.  The black
points are photographic Johnson $V$ data from Schombert (1986), corrected
for a $V-J=2.57$ color.  The
correspondence between the two profiles is excellent considering the difference
in wavelength, detector type (photographic versus electronic) and time (15
years).  There appears to be a slight blue gradient consistent with
expected $V-J$ metallicity gradients.  The data for the current project is
only one magnitude brighter than the $V$ data limits (sky noise limited).
}
\end{figure}

Of the 15 galaxies in common with the Schombert (1986) sample, 80\% were in
good agreement with the new $J$ data, meaning the slope of the profiles
agree within the errors and the photometric offset corresponds to within
20\% of the $V-J$ color (see a later paper for more detailed discussion of
color profiles).  Disagreements were typically due to strong color
gradients or poor calibration in the $V$ data (based on recent $V$ aperture
measurements).  While this does not prove absolute repeatability of our
current dataset, it does give us some confidence in the reliability of the
profiles.  In addition, it will be argued in a later paper, that the
largest source of uncertainty in determining the structure of galaxies is
not the errors in the photometry, but rather the interpretation of the
profiles.

\subsection{PSF Effects}

The inner regions of any galaxy surface brightness profile will be
distorted by the inherent limitations of atmospheric and detector
resolution.  This distortion is represented by the point source function
(PSF), and the effects of the PSF on the surface brightness profiles of
galaxies is well studied by Saglia \etal (1993).  The 2MASS PSF 
is also well documented for 2MASS images (Jarrett \etal 2000), and is 
gaussian in shape with typical FWHM's ranging from 2.5 to 3 arcsecs.

\begin{figure}[!ht]
\centering
\includegraphics[scale=0.8]{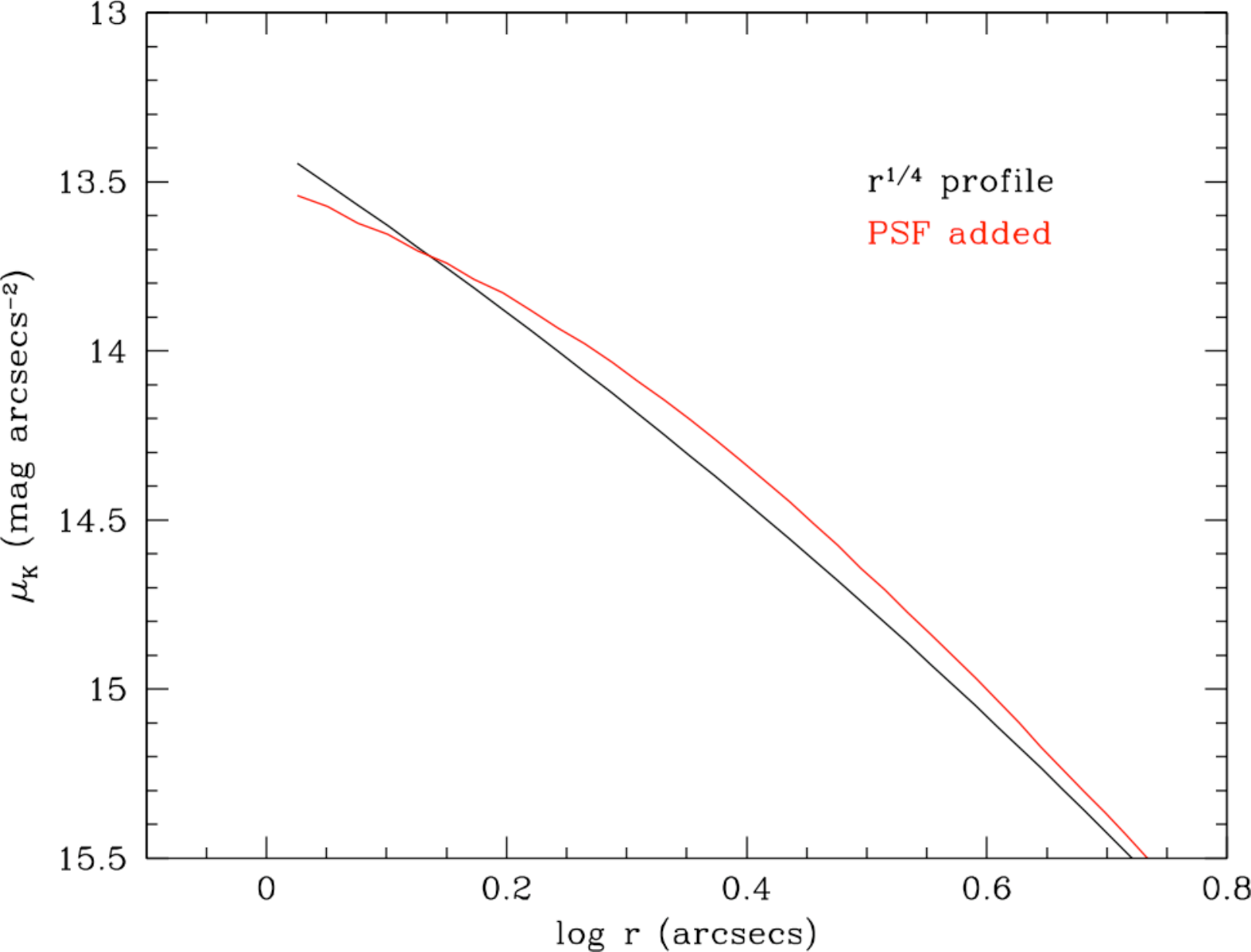}
\caption{\small A pure r$^{1/4}$ galaxy profile (black) is convolved with a
typical 2MASS PSF using a FWHM of 2.5 arcsecs.  The simulated galaxy
profile uses the mean effective surface brightness ($\mu_e$=17.0 $K$ mag
arcsecs$^{-2}$) and mean effective radius ($r_e$ = 10 arcsecs) of our
elliptical sample.  The resulting PSF adjusted profile shows distortion
from the original galaxy profile out to 5 arcsecs.}
\end{figure}

The PSF of an image is produced when atmospheric and detector distortion
moves core photons to larger radii.  The resulting surface brightness
profile is slightly dimmer at the core, and slightly brighter at the wings
of the PSF.  For the 2MASS images, the PSF is primarily driven by the large
detector pixels used (2 arcsecs), but some resolution is recovered by a
dithering observing strategy.  For a majority of the 2MASS images, the mean
2MASS PSF is determined to have a FWHM of 2.5 arcsecs, although when
atmospheric seeing degrades below this level, the FWHM increases (although
2MASS restricted data acquisition during poor seeing conditions). For our
experiments we have assumed a FWHM of 2.5 arcsecs.

To generate a 2MASS PSF, we have taken the gaussian description from Saglia
\etal (1993, see their Figure 6) where the FWHM/2$r_e$ ratio is the model
variable ($r_e$ is the effective scalelength for an assumed r$^{1/4}$
profile shape, see Appendix C for how these simulations are generated).  As
a test simulation, we have adopted a pure r$^{1/4}$ shape using mean
parameter values from our elliptical sample ($r_e$ = 8.5 arcsecs, $\mu_e$ =
16.7 $K$ mag arcsecs$^{-2}$).  Note that this is a poor description of the
outer regions of galaxies, but an adequate one for the core regions
(Schombert 1987).  For a FWHM of 2.5, this corresponds to a Saglia PSF of
0.15.  The resulting profile is shown in Figure 5.
Note that the 2MASS PSF distorts a galaxy's surface brightness profile out
to at least a radius of 5 arcsecs.

\begin{figure}[!ht]
\centering
\includegraphics[scale=0.8,angle=0]{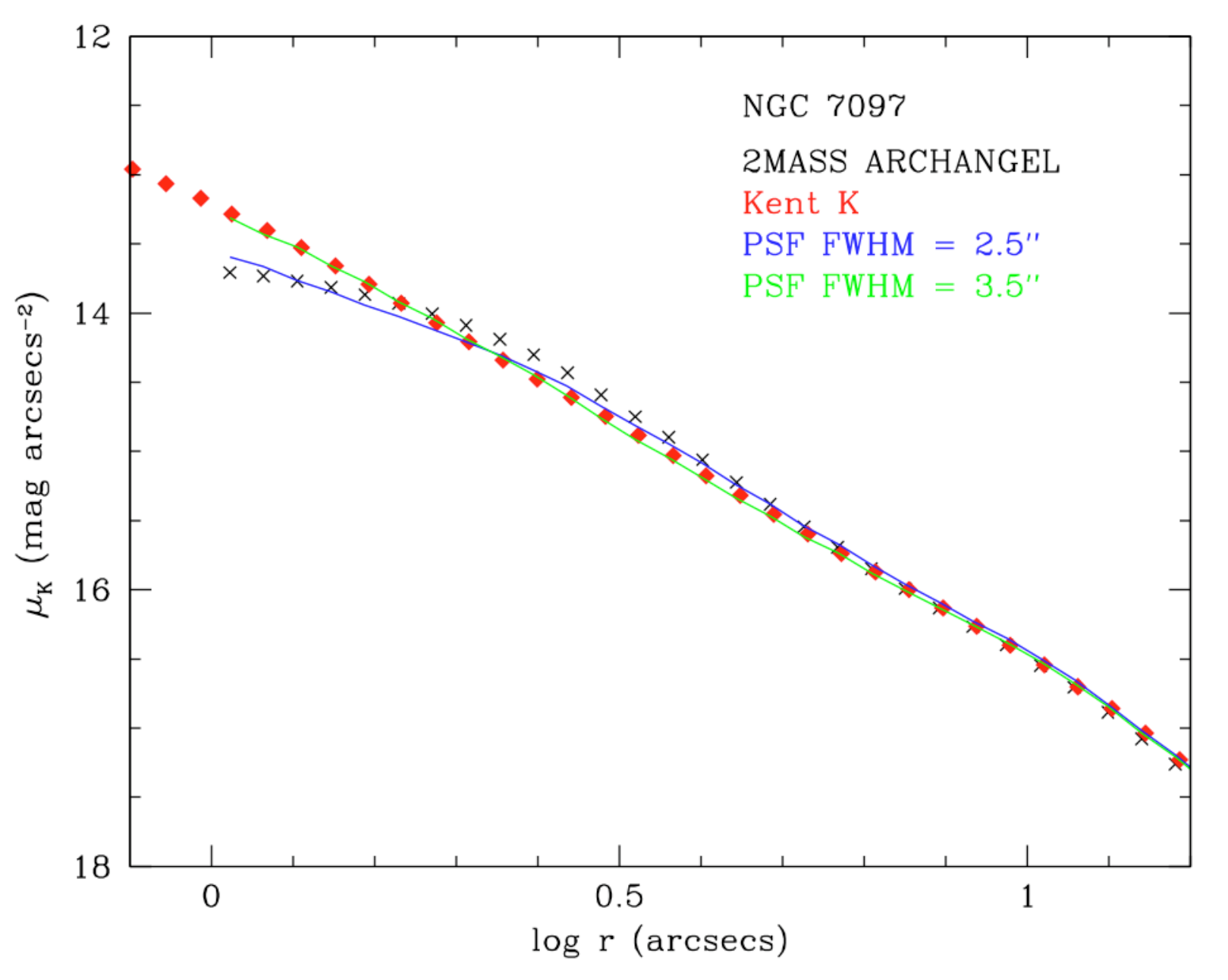}
\caption{\small An example of PSF corrected 2MASS profiles for the
elliptical NGC 7097.  The 2MASS profile is shown as black symbols, high
resolution (0.2 arcsecs pixles) $K$ band imaging from Kent (2012) is shown
as red symbols.  The 2MASS profile is corrected for a 2.5 (blue) and 3.5
(green) arcsec FWHM.  While typical 2MASS PSF's are quoted at 2.5 arcsecs,
the 3.5 arcsecs PSF recovers the higher resolution profile.  Again, 5
arcsecs appears to be the outer radius for PSF distortions to 2MASS surface
brightness profiles.
}
\end{figure}

As an additional test of the effects of the PSF, we compare the 2MASS
profile for NGC 7097 with a deep $K$ image from Kent (2012) shown in Figure
6.  The profiles agree fainter than 16 $K$ mag
arcsecs$^{-2}$, but the effects of the PSF distortion are visible in the
2MASS data at brighter surface brightnesses (the Kent data was taking under
sub-arcsec seeing with 0.2 arcsecs pixels).  We correct the 2MASS profile
with two FWHM's of 2.5 (blue curve) and 3.5 (green curve) arcsecs.  While a
FWHM of 2.5 arcsecs is typical for 2MASS images, the 3.5 arcsecs PSF is a
better fit producing agreement with pointed observations down to a radius
of 1.5 arcsecs.  Information below that radius point is difficult to
recover without some assumptions to the shape of the core region of a
galaxy, a problematic approach given the core/cusp dilemma for ellipticals.

The shape of the PSF from Figure 6 is salient to fitting
surface brightness profiles.  One could attempt to correct a galaxy's
surface brightness profile, thus, gaining a few arcsecs of resolution at
the core.  However, this makes the assumption of an underlying r$^{1/4}$
profile which, while a reasonable assumption for early-type galaxies,
defeats the purpose of actually measuring the structure of galaxies.  The
more honest evaluation of the profiles would be to only fit 2MASS galaxy
profiles outside the 5 arcsec radius and this study will apply this 5
arcsec inner limit, even when correction to an r$^{1/4}$ profile seems
appropriate.

\subsection{Error Analysis}

The source of error in surface brightness profiles is threefold: 1) error
in the detector and photometric calibrations, 2) RMS noise around the best
fit ellipse (this includes error due to deviations from an elliptical
shape for a particular isophote) and 3) error in the value of sky.
Numerical experiments with ellipse fitting on both symmetric and irregular
galaxies has shown that the surface photometric values are very robust to
variations in the ellipse parameters, contributing less than 2\% to the
photometric noise (Schombert 1986).  Cleaning too many pixels can have a
negative effect on the inner regions of a galaxy, but is negligible in the
outer regions where the S/N is the lowest.

Addressing each of these sources in order, the detector and photometric
calibration for pointed observations can be problematic.  However, the
2MASS sky survey calibrates many times a night and the characteristics of
the detectors are extremely well known.  The observing strategy eliminates
most pixel-to-pixel errors.  Given the project reports on the photometric
calibration (Skrutskie \etal 2006), we assume the calibration error is
negligible.

The RMS noise around each ellipse is calculable and stored in the raw data
files.  For most regular shaped galaxies, the S/N does not decrease below
five until the surface brightness reaches one magnitude below sky.  Until
this point, the RMS errors are strictly photon count limited and decrease
in importance as the number of pixels in the ellipse increases.  For the
outer isophotes, where the number of pixels is large, the calculated RMS on
the mean intensity produces an artificially low estimate of the error.
For, while the value of the mean intensity is more accurate because of
increases in N, its absolute value is more uncertain due to sky error.

\begin{figure}[!ht]
\centering
\includegraphics[scale=0.7,angle=0]{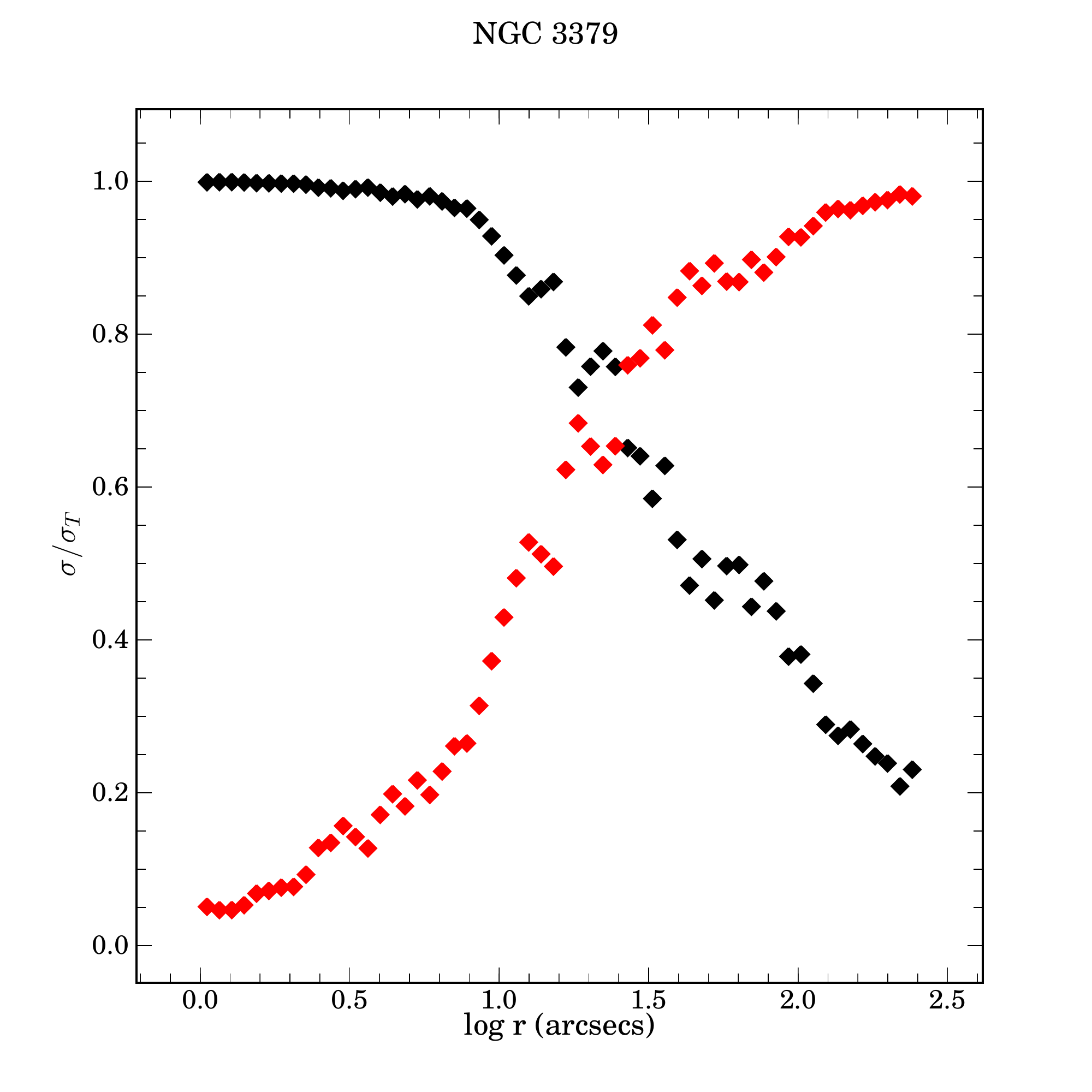}
\caption{\small A comparison of the error contribution from RMS noise
around each ellipse (black) and the error in sky determination (red).  In
the inner regions, where the surface brightness is high, RMS noise
dominates the error budget.  But, only at 1/10th the total radius of the galaxy,
the sky error begins to dominate.  Due to the large number of pixels in the
outer envelope, over 80\% of the luminosity of a galaxy has its uncertainly
fixed by knowledge of the proper sky value.
}
\end{figure}

Of greater concern is the exact knowledge of the sky value.  Figure
7 displays the error budget for a typical elliptical
comparing the error due to RMS noise and the error associated with the
knowledge of the correct sky value.  Typically, at the point where the
effective radius is reached, the sky error dominates the surface brightness
error budget.  Since a majority of a galaxy's light is beyond this radius,
then any parameter that involves the total luminosity of the galaxy is
limited by the correct knowledge of the sky value.

\begin{figure}[!ht]
\centering
\includegraphics[scale=0.7,angle=0]{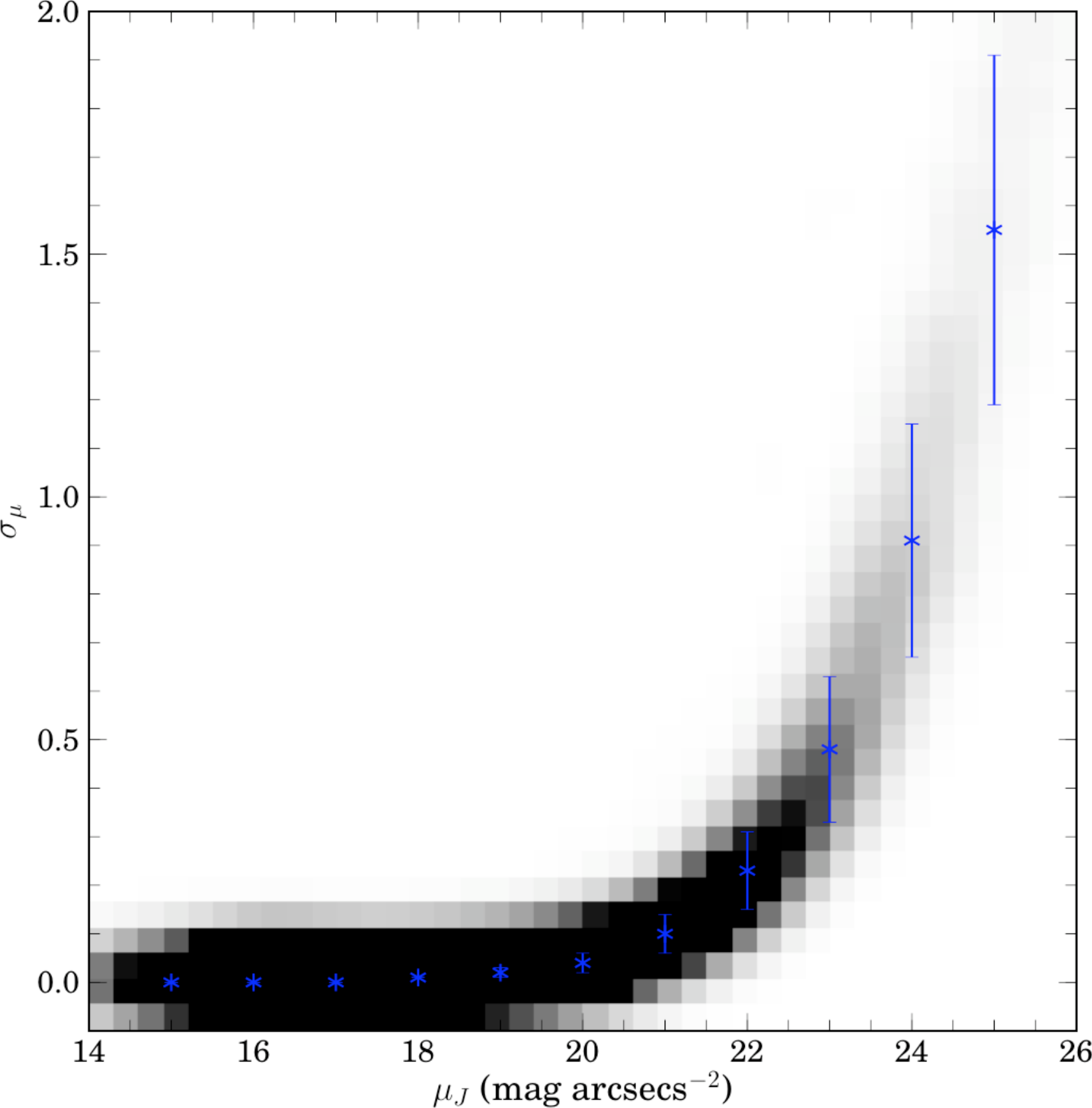}
\caption{\small The number density distribution of surface brightness error for all
early-type galaxies in the 2MASS sample.  As sky error dominates the error
budget at low surface brightnesses and, since 2MASS imaged the sky under similar
detector and sky conditions, the range of errors is well defined.  The
greyscale shows the number density of all surface brightness points for
the early-type galaxy sample.  The blue points are the average and standard
deviation of that data.  Errors exceed one mag arcsecs$^{-2}$ below
24.5 $J$ mag arcsecs$^{-2}$.
}
\end{figure}

With respect to the total error values for the surface brightness profiles,
Figure 8 displays all the error values assigned to each
surface brightness point in our elliptical sample.  These values are the
quadrature sum of the RMS noise and the error on the sky value.
Given the narrow range of sky variation and common detector, the errors are
remarkable consistent with surface brightness.  Basically, the errors grow
larger than the meaning of the observations below 25 $J$ mag arcsecs$^{-2}$
(which corresponds to roughly 27 $V$ mag arcsecs$^{-2}$).  Thus, the data
presented here is similar to deep optical samples in the past (Schombert
1986).

\subsection{Fitting Functions}

The last stage in the data reduction pipeline is the fitting of the
uncorrected surface brightness profiles to various fitting functions.  The
choice to fit to the uncorrected magnitudes and radii in arcsecs merely
allows other researchers to apply their own corrections.  As extinction and
distance corrections are in the data files, this becomes a stylistic point
rather than a critical part of the reduction process.  All the fitting
variables can be corrected for extinction and distance without changing the
quality of the fits themselves.

The profiles in this study are fit to the three most popular fitting
functions 1) de Vaucouleurs r$^{1/4}$ (de Vaucouleurs 1959), 2) S\'{e}rsic
(Graham \& Driver 2005) and 3) r$^{1/4}$ bulge plus exponential disk
(bulge+disk, Freeman 1970, Kent 1985).  While there are other fitting
functions in the literature (Oemler 1976), these three are mathematically
equivalent to any other function.  The r$^{1/4}$ law has two parameters
(effective surface brightness, $\mu_e$, and effective radius, $r_e$).  The
S\'{e}rsic function has three parameters ($\mu_e$, $r_e$ and the power law
index, $n$).  The bulge+disk function has four parameters (bulge $\mu_e$,
$r_e$, central disk surface brightness, $\mu_o$, and disk scalelength,
$\alpha$, sometimes called $h$ in the literature).  Note that a pure disk
profile is simply a bulge+disk fit with a zero sized bulge.

\begin{figure}[!ht]
\centering
\includegraphics[scale=0.7,angle=0]{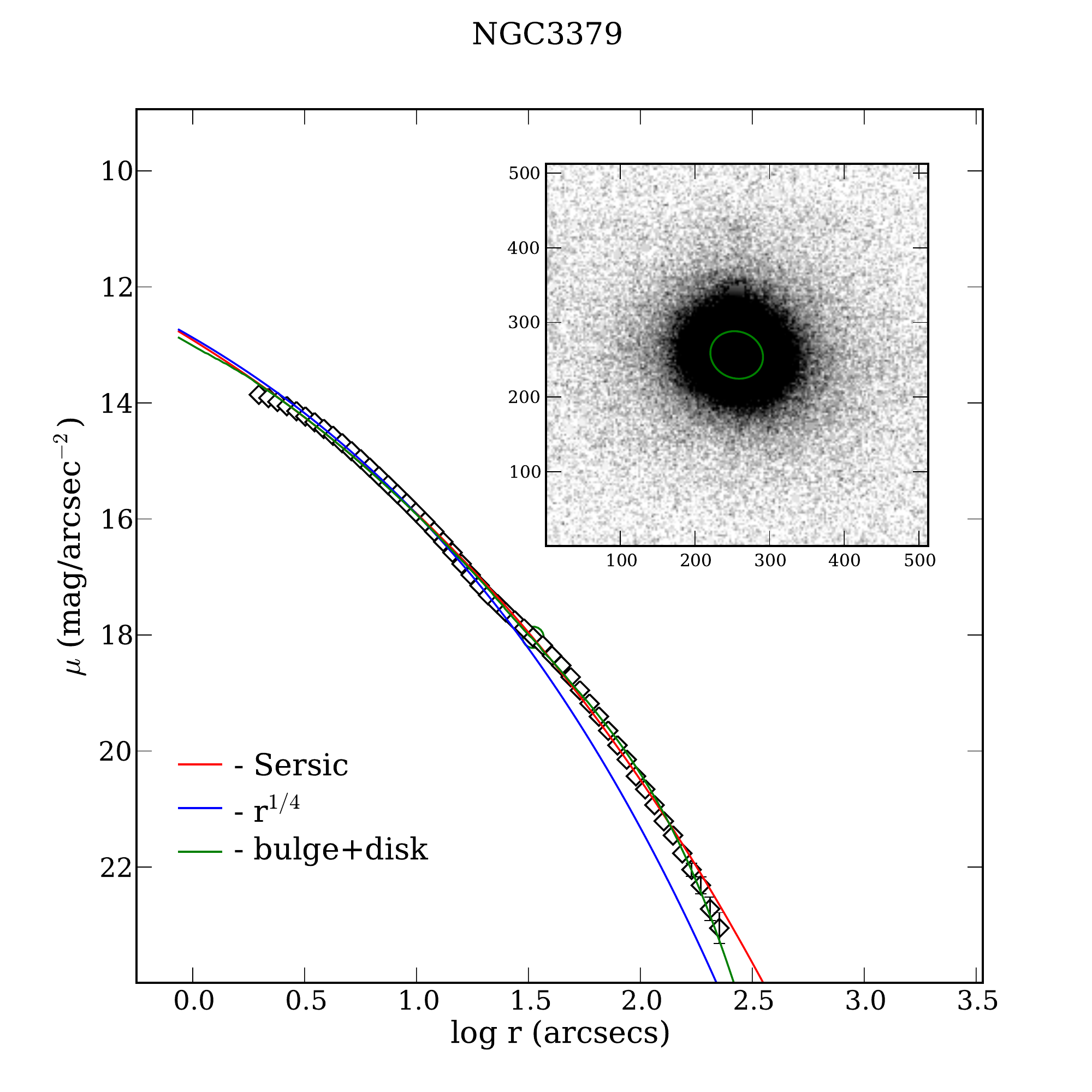}
\caption{\small The surface brightness profile for NGC 3379, a standard
elliptical galaxy.  The best fit of the three most popular fitting
functions are shown.  The S\'{e}rsic function has the formal best fit.  The
green circle indicates where an inflection point exists for most bright
ellipticals (that radius is also indicated in the greyscale image).  The
r$^{1/4}$ law is only fit to the middle portion of the profile, galaxies
are not r$^{1/4}$ in their shape for a range of surface brightness and
luminosity (Schombert 1987).
}
\end{figure}

This project's procedures for applying each function deviated from accepted
practices in the literature.  The r$^{1/4}$ law is fit by plotting the data
in surface brightness vs r$^{1/4}$ space and isolating the region of the
galaxy that displays a straight line for fitting, ignoring other data.
Typically, the r$^{1/4}$ region of a galaxy's surface brightness profile is
the middle portion of a galaxy's profile (between 14 and 17 $J$ mag
arcsecs$^{-2}$), where the inner regions display the core/cusp dilemma and
the outer regions develop curvature in a luminosity dependent fashion
(Schombert 1986).  While this is a subjective fitting procedure (visual
inspection determines the region to fit), fits made to the entire profile
leads to erroneous parameters since galaxies are simply not r$^{1/4}$ in
their shape over all regions and luminosities.

The S\'{e}rsic function is fit over the region of a galaxy profile outside
the seeing effected core and stopping where the error in the surface
brightness exceeds one mag arcsecs$^{-2}$.  This procedure is automatic
and, of the three fitting functions, is the most objective.  In general, a
S\'{e}rsic fit is superior to an r$^{1/4}$ fit simply because the $n$ index
adds an additional free parameter that primarily branches the non-r$^{1/4}$
portion of the outer envelope of a galaxy.  We will explore this in greater
detail in a later paper.

The bulge+disk fitting follows the prescription given by Schombert \&
Bothun (1987).  First, a linear portion of a galaxy's outer region
(i.e., the disk) is visually located and fit to an exponential (a straight
line in mag arcsecs$^{-2}$ versus radius space).  Holding the slope of the
disk fit constant (but allowing the central surface brightness to vary) a
r$^{-1/4}$ fit is applied to the inner regions (the bulge).  The three
parameter bulge+disk fit is recorded, than the constraint on the disk slope
is released and a four parameter fit is made.  Strong changes in the disk
slope between the first and second fits signals a galaxy which did not have
a strong disk component at the start, but rather is dominated by a
power-law bulge.

A technical note, fits made with the bulge+disk function assume rotation
generated circular symmetry to the galaxy, so the major axis of each
ellipse is used on the assumption that the minor axis length is due to
orientation on the sky.  Fits using the r$^{1/4}$ or S\'{e}rsic functions
do not assume circular symmetry and, thus, the generalized radius
($ab^{-1/2}$) is used.  Ellipticals are not typically oblate, thus the
bulge+disk fit simply becomes an open four parameter polynomial.

\begin{figure}[!ht]
\centering
\includegraphics[scale=0.7,angle=0]{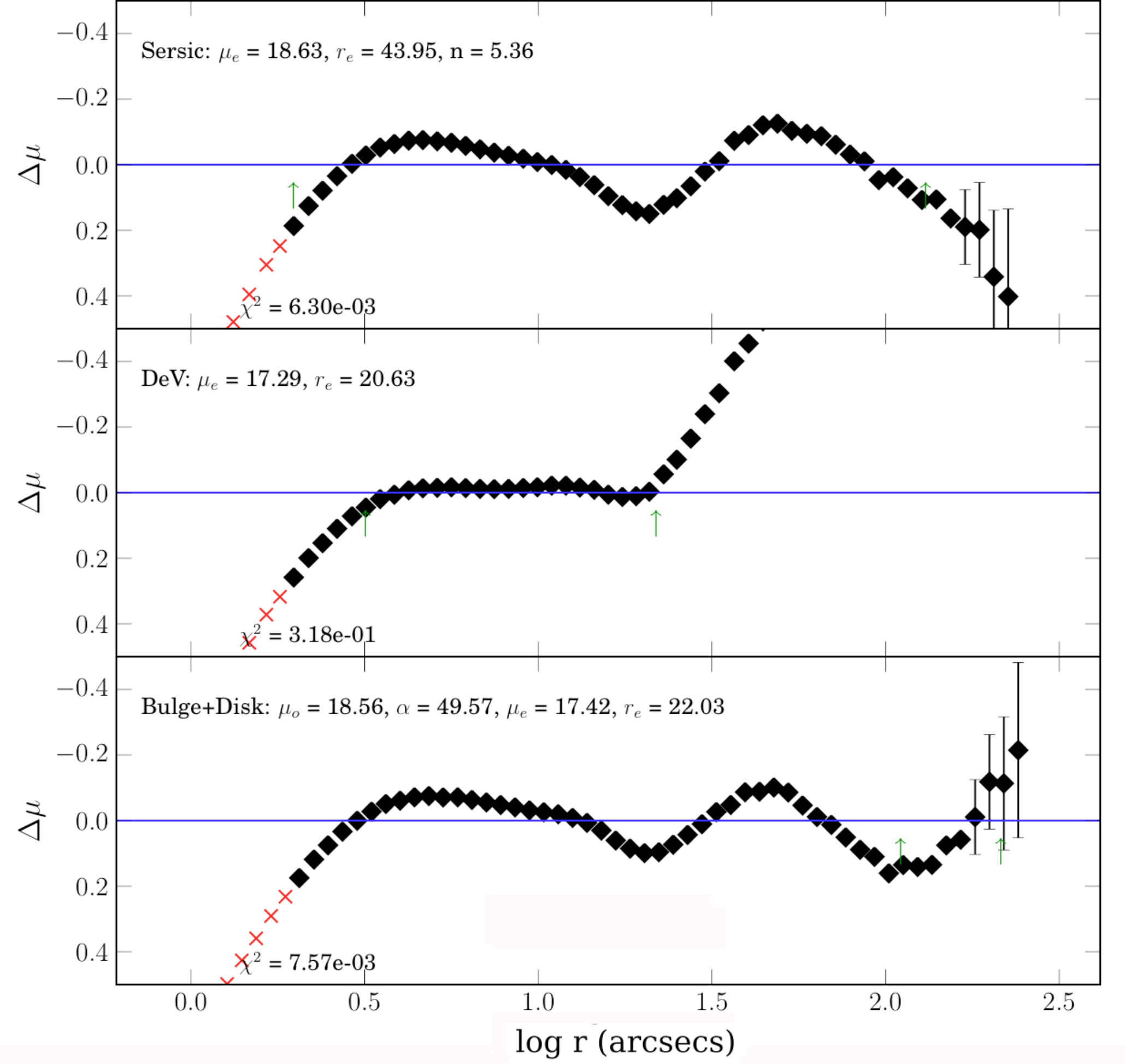}
\caption{\small The residuals from the three fitting functions for NGC
3379.  The curvature in the residuals is correlated with galaxy luminosity
(mass) indicating that fitting functions fail to describe the structure of
galaxies in a consistent fashion.
}
\end{figure}

An example of all three fitting functions is shown in Figure
9, a plot of the $J$ surface brightness profile for NGC
3379.  The quality of each fit is evaluated using a simplified $\chi^2$
estimator where all the datapoints are equally weighted and the $\chi^2$
simply becomes the sum of the square of the differences.  Weighting the
data by its photometric error is the normal procedure, but this gives too
much weight to the inner isophote datapoints since their RMS errors are
very small and their larger number (due to the smaller ellipse annuli)
overwhelms the data in the outer regions.  This produces erroneous fits,
particular since the outer regions are often the most interesting for
determining global structure parameters (such as half-light radii).  This
is particularly problematic for galaxies with bulge and disk components,
where the brighter and more compact bulge component dominates the fit over
a few outer disk points.  The $\chi^2$ values shown in the Figures are
unweighted for comparison between profiles.  A later paper will detail the
different fitting strategies adopted for different morphological types.

Of the three fitting functions, the bulge+disk function provides the lowest
formal $\chi^2$ value, despite the fact there is no obvious evidence of a
disk.  This is simple due to the fact that four parameters provides more
flexibility to the fitting, resulting in a technically better fit.  The
S\'{e}rsic has the lowest formal $\chi^2$, using three free parameters.
The r$^{1/4}$ law is only fit to the middle portion of the profile and
fails in the outer envelope.  The green circle in Figure 9
marks a common inflection point for bright ellipticals.  Typically where
this inflection point occurs is the point where a bulge+disk fit separates
the disk component.

A better method to display this information is shown in Figure
10 where the residuals to the various fitting functions are
shown.  While the formal fit is best for the S\'{e}rsic function, we will
show in Paper II that each function has specific deviations from a best fit
that are correlated with luminosity.  Our conclusion is that all fitting
functions are simply computational French curves that only contain
information on structure as constrained by the procedure for applying them.
We will discuss the usefulness of fitting functions in greater detail for
each morphological type in later papers.

\subsection{Data Storage/Access}

Data storage and presentation for a image reduction project of this type
entails many complications.  In the past, researchers would simply publish
the resulting surface brightness profiles as luminosity versus radius.
However, this habit of only presenting the finished product has two
disadvantages.  First, while reduced data is the goal of any imaging
project, in fact, there is a great deal of information contained in the
processing files.  For example, the Fourier quotients on the ellipses
contains information on the shape of the isophotes as they deviate from a
perfect ellipse (i.e., disk versus boxy).  Interpretation of the surface
brightness profiles is critically dependent on that information.
Luminosity by apertures, color gradients, spatial anomalies are also
contained in the processing files.

Second, there is a level of transparency to the reduction process by
presenting all the data including the raw and processed data.
Repeatability is a key component to the scientific process.  Presenting all
the processed data, and the actual software used to process the frames,
is more than just a statement of the honesty of a dataset, but also key to
understanding the meaning of the final numbers.

To this end, this project maintains all the final and processed data in XML
files linked to the raw image frames.  In these files are all the
parameters used during the processing and calibration of the images.  For
example, if a user selected region of the surface brightness profile is
used for fitting, the fit and the user selected limits are recorded.
Cosmological corrections, such as galactic extinction and distance
(obtained from NED) are also stored in these files.

In addition, the scripts used to process the data are presented with the
data.  Rather than publishing appendices describing the algorithmic
procedures to each reduction step, these scripts, loaded with comments,
guide the user in both the computational and astrophysical data analysis
steps.  These scripts are Python code, which often call C++ or IRAF
subroutines.  Python is ideal for this type of pipeline processing as it
can handle numerical and text decision processing as well as file and
directory instructions.  With the addition of the PyFITS module from the
PyRAF project, these scripts can now also access information with the
images themselves (e.g., headers and pixel values).  A cookbook example of
the processing is available at the data website (http://abyss.uoregon.edu/
$\sim$js/sfb) although with descriptions of the data values and the raw and
reduced images.

\section{Summary}

This paper presents the techniques and philosophy for a large scale, galaxy
surface photometry project using 2MASS imaging data.  Our ultimate goal is
to investigate the surface brightness profiles across all morphological
types providing a comprehensive view of the structure of galaxies.  We
summarized the key points of this first paper of our series as the
following:

\begin{description}

\item{(1)} The 2MASS all-sky survey presents an ideal database in which to
study galaxy structure.  For imaging in the near-IR emphasizes the primary
baryonic component of galaxies, stars.  The observing technique used by the
2MASS project produces extremely flat and well calibrated images, the two
primary sources of error in pointed observations.

\item{(2)} We have developed unique network tools to automatically extract
and assemble regions from the 2MASS image server based on input catalogs.
For our goals, we have selected the Revised Shapley-Ames (a luminosity
limited) and the Uppsala (a diameter limited) galaxy catalogs from which to
extract our galaxies.

\item{(3)} We have made no attempt to be complete in our sample in terms of
galaxy luminosity or size nor galactic latitude, although we have attempted
to reduce every large galaxy in the sky in order to maximize our samples
range of galaxy characteristics.  We have eliminated all galaxies in our
initial catalog selection that had companions, or nearby bright stars,
which would complicate our analysis.  Our goal was to obtain as many
isolated galaxies for study per morphological bin as was possible with the
limited depth of the 2MASS survey.

\item{(4)} Data reduction used the ARCHANGEL galaxy photometry package
(Schombert 2007).  Particular attention was given to star removal and sky
determination (as these are the two main contributors to error in the
resulting surface brightness profiles).

\item{(5)} Repeatability is supported by comparison with surface brightness
profiles in the literature.  The error budget is completely dominated by
errors in the sky value.

\item{(6)} Fitting functions, and our procedures for using them, are
outlined.  We use only three of the more common fitting functions
(r$^{1/4}$, S\'{e}rsic and r$^{1/4}$ bulge plus exponential disk).

\item{(7)} We present our method of data storage in an attempt to open the
access to all levels of the data product as a great deal of structural
information is found beyond simple fits to the final surface brightness
profiles.

\end{description}

Four appendices are attached to this paper that A) outlines the problems in
the 2MASS Extended Source Catalog (XSC) with respect to total magnitudes
and surface brightness, B) provides an example our our asymptotic total
magnitude procedure, C) outline the PSF results and procedures, and D) list
the variables found in the XML data files for each galaxy.

\noindent Acknowledgements: Most of the comparative values were extracted from NED
(NASA's Extragalactic Database) using new network tools.  The quick access
to difference galaxy catalogs on one site made this project doable in
reasonable timescales.  The model for future science is not faster cycles,
but faster and clearer access.  The software for this project was supported
by NASA's AISR and ADP programs.

\appendix
\section{Comparison to 2MASS Total Magnitudes}

During the initial stages of this project a comparison between total
magnitudes determined for the elliptical sample of this project and the
values provided by 2MASS XSC revealed a large discrepancy.  This appendix
describes the analysis that lead to the conclusion that the 2MASS XSC was
consistently underestimating the total magnitude values due to a systematic
error in their sky measurements.  Note: this appendix was released to arXiv
as Schombert (2011), "Systematic Bias in 2MASS Galaxy Photometry".

\begin{figure}[!ht]
\centering
\includegraphics[scale=0.7,angle=0]{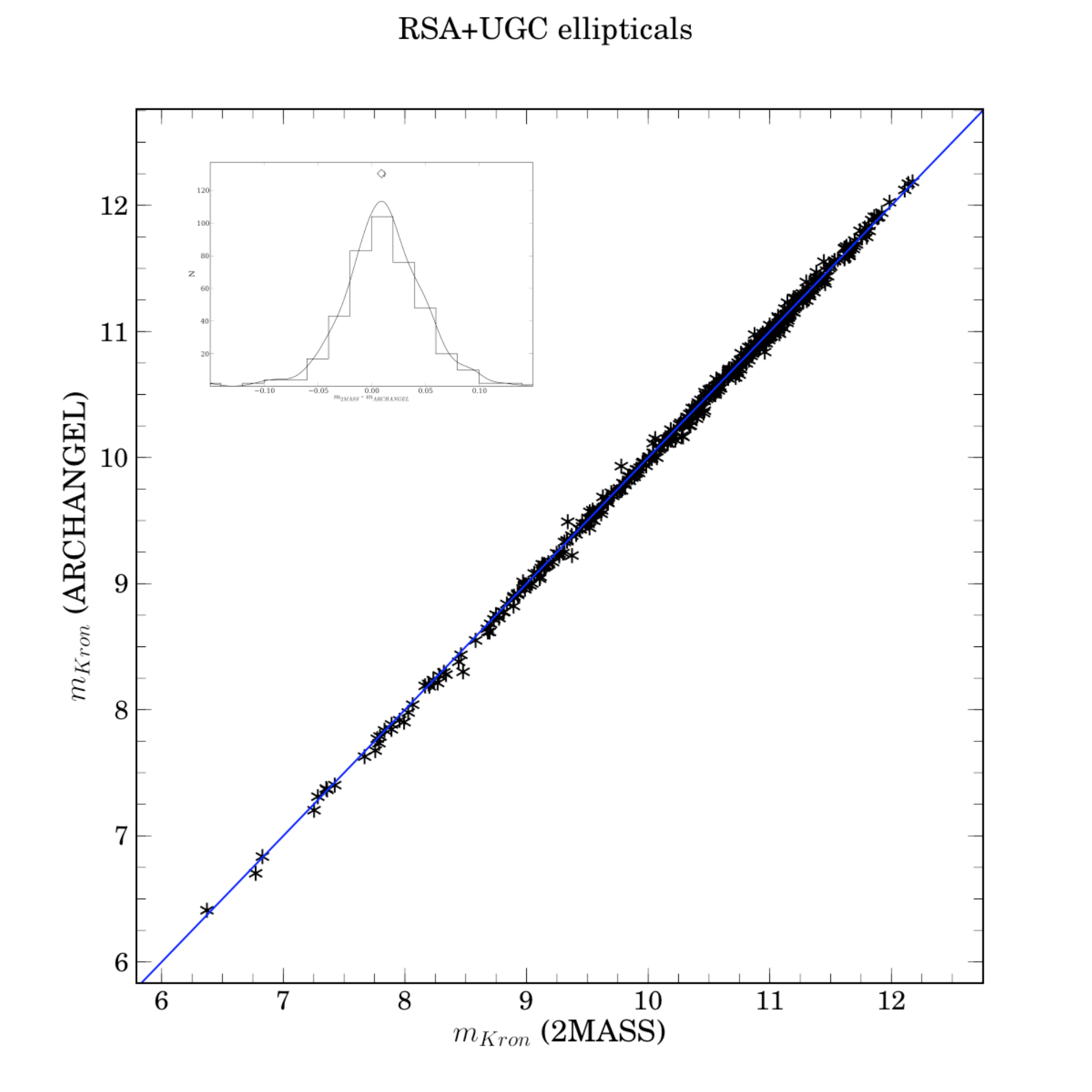}
\caption{\small Comparison of 2MASS $J$ Kron apparent magnitudes for 421 ellipticals
on our galaxy structure survey (Schombert \& Smith 2011).  The blue line is
the one-to-one equivalence line.  The agreement is excellent as we use the
aperture sizes and orientations given by the 2MASS project.  The inset
histogram is the difference in magnitudes, the mean difference is 0.01 mags
(our magnitudes are slightly fainter because of our pipeline reduction
procedures that subtract stars and replaces their pixels with interpolated
galaxy flux).
}
\end{figure}

A comparison sample of elliptical galaxies was selected from the Revised
Shapley-Ames (RSA) and Uppsala Galaxy Catalogs (UGC) in order to cover a
magnitude and angular limited sample with sufficient S/N in the 2MASS image
library.  The only other criteria was that the galaxies to be studied be
free of nearby companions or bright stars which might disturb the analysis
of the isophotes to faint luminosity levels.  The comparison sample
contained 421 galaxies all classed 'E' by both catalogs.

Images from 2MASS for regions around all the galaxies in the sample were
downloaded from 2MASS's Interactive Image Service.  These sky images were
flattened and cleaned by the 2MASS project and contained all the
information needed to produce calibrated photometry.  The images were
analyzed as described in \S 3, thus, the only difference in the final
results is the analysis method, not the data themselves.

\subsection{2MASS Repeatability}

The first step, once surface photometry reduction was completed, was to
compare our photometric and structural values with those extracted by
2MASS.  Metric magnitudes are the simplest for comparison, and the 2MASS
project provides magnitudes through various aperture sizes (e.g. 14 arcsecs
apertures are found in NED).  The 2MASS project also provides Kron
magnitudes, where Kron magnitudes are isophotal magnitudes measuring a
galaxy's light through an elliptical aperture whose size is defined by the
20 $K$ mag arcsecs$^{-2}$ surface brightness level.  These magnitudes
contain less intrinsic error than metric magnitudes as the Kron apertures
follow the shape of the galaxy and maximizes the galaxy flux to sky ratio.
NED provides those magnitudes and the aperture sizes for all the galaxies
in our sample.  A comparison between our Kron magnitudes (using 2MASS's aperture
sizes) with their Kron magnitudes is shown in Figure 11.

\begin{figure}[!ht]
\centering
\includegraphics[scale=0.7,angle=0]{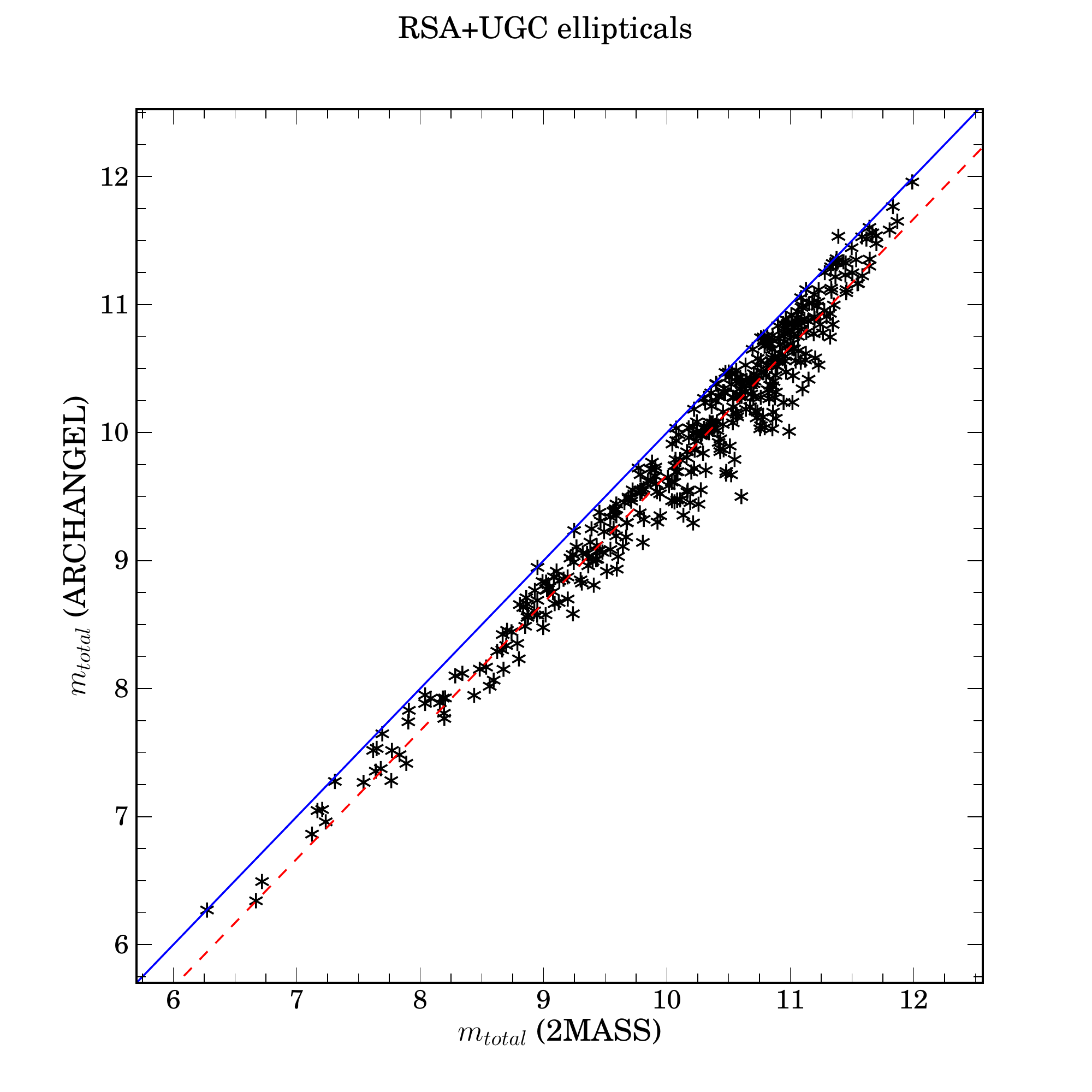}
\caption{\small Comparison to total magnitudes ($J$ band) from the 2MASS
Extended Source Catalog with our photometry from the 2MASS raw images.  The
blue line represents one-ton-one correspondence, the red line is a linear
fit with a slope of 1.001.  The 2MASS total
magnitudes are 0.33 mags fainter than our calculated total magnitudes.
This represents an error ranging from 10 to 40\% in total luminosity.
}
\end{figure}

As can be seen from Figure 11, the agreement between our Kron
magnitudes and 2MASS values is excellent, meaning that we can reproduce the
same fluxes as the 2MASS project using the same apertures off of 2MASS
provided images and calibration.  There is a slight offset (0.01 mags) such
that our magnitudes are slightly fainter than 2MASS (see inset histogram).
This is probably due to the fact that our reduction procedure subtracts stars and replaces
the masked pixels with interpolated galaxy flux which, on average, would
lower the aperture flux.  We note that our values for the Kron ellipses
were significantly larger than 2MASS's estimates.

\subsection{Problems with 2MASS Total Magnitudes}

The next comparison is with our total magnitudes and 2MASS's total
magnitudes.  This comparison can be found in Figure 12 and,
as is visible in the Figure, a significant difference is found between our
calculated total magnitudes and the values presented for the same galaxies
by the 2MASS project.  In general, our total magnitudes are 10 to 40\%
brighter than the 2MASS total luminosities.

This discrepancy in total luminosities is especially puzzling since our
reduction methods can reproduce 2MASS aperture and Kron magnitudes (see
Figure 12).  This would indicate that the images provided by
the 2MASS project are reliable and the calibration is correct.  The
difference must lie in the reduction procedures to determine total
magnitudes.

Total magnitudes determined by the 2MASS project use an aperture magnitude
that is four scalelengths in radius, where the scalelength is determined
S\'{e}rsic fits to their surface brightness profiles.  Our project
determines total magnitudes through asymptotic fits to the curve of growth,
where we increase the accuracy of the outer isophotes by using the mean
intensities give by the surface brightness profiles (see Appendix B).

\begin{figure}[!ht]
\centering
\includegraphics[scale=0.8]{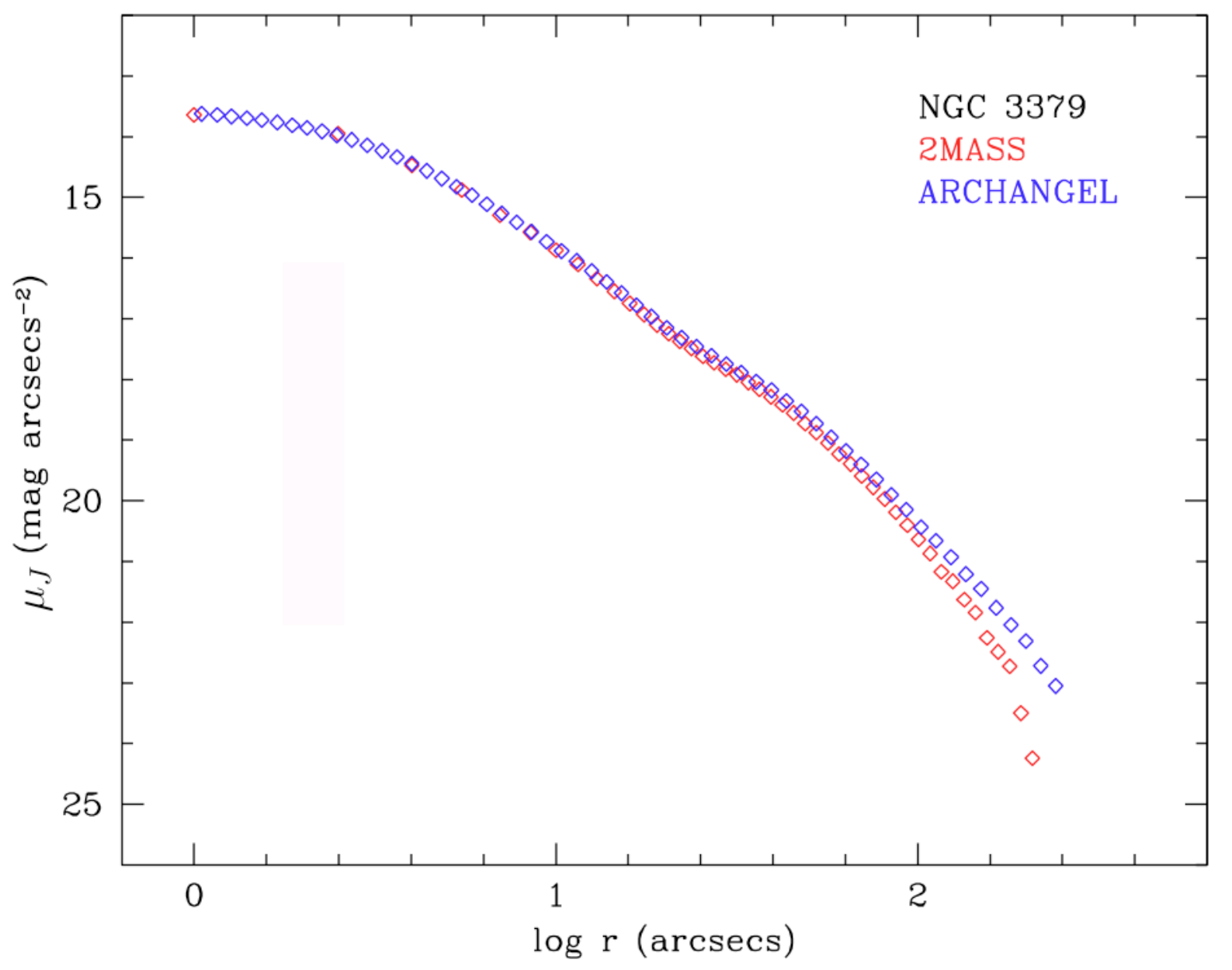}
\caption{\small A comparison of the $J$ surface brightness profile
presented by the 2MASS project (Jarrett \etal 2003) and the profile reduced by
our software package (ARCHANGEL).  The photometry agrees at high surface
brightnesses, but begins to disagree below 18 $J$ mag arcsecs$^{-2}$.  As
discussed in the text, the difference can not be explained by poor ellipse
fitting, calibration error or an improper sky value.
}
\end{figure}

The key difference in our photometric methodology lies in the determination
and use of the galaxy's surface brightness profile to determine the
aperture.  Therefore, an error between the surface brightness profiles
deduced by our procedures and the ones extracted by the 2MASS project must
be the source the magnitude offset.  To explore this hypothesis, we compare
the procedures used by 2MASS and our procedures in the next section.

\subsection{Surface Photometry Comparison}

The 2MASS project also published surface brightness profiles for 100 large
galaxies (Jarrett \etal 2003), 31 of them in common with our elliptical
sample.  Agreement between our surface brightness profiles and the 2MASS
project's profiles is less than adequate.  An example is found in Figure
13, the surface brightness profiles of NGC 3379 from Jarrett
\etal and our study.  The difference between the profiles is extreme at
large radii, well beyond expectations from the RMS errors in the data.

\begin{figure}[!ht]
\centering
\includegraphics[scale=0.8]{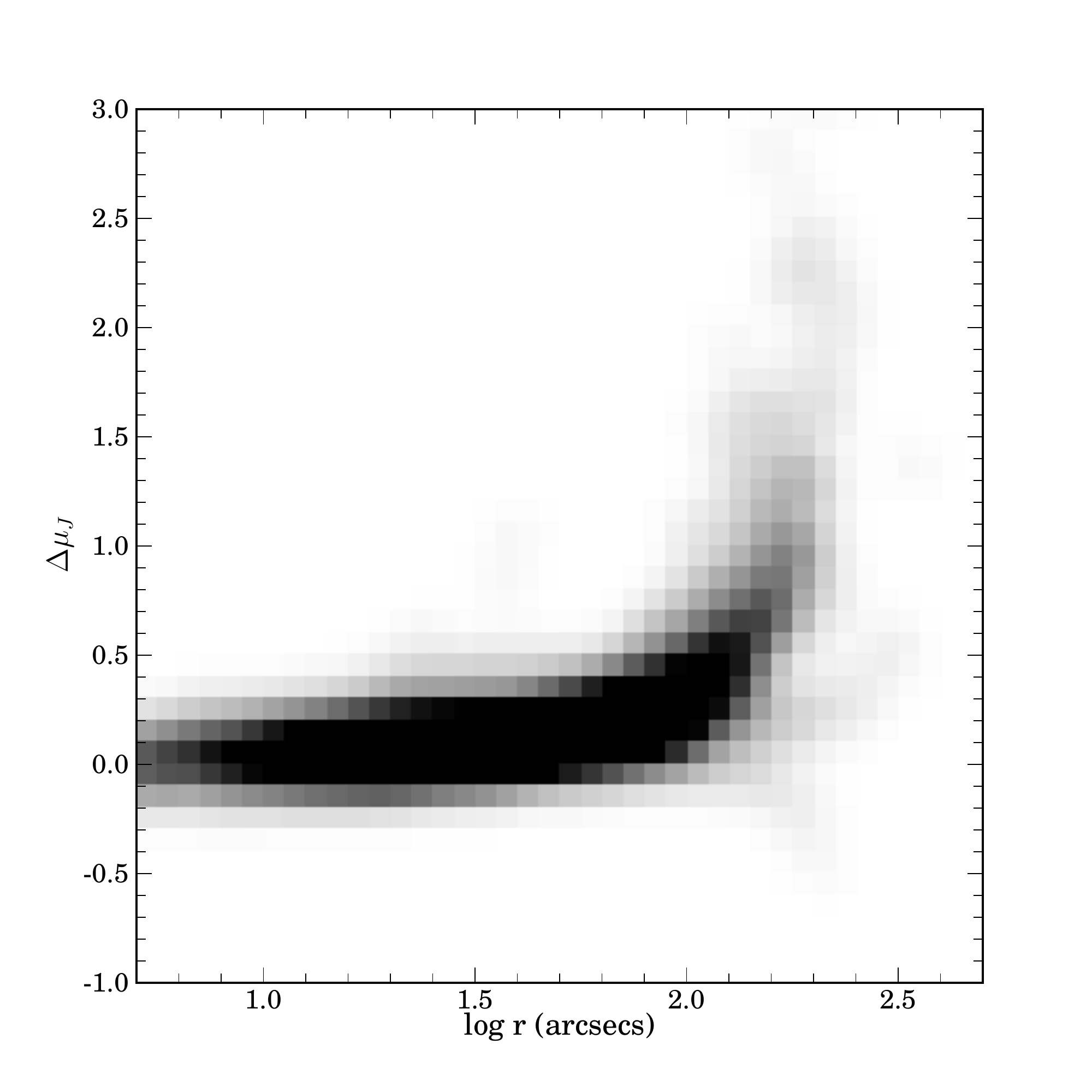}
\caption{\small A density plot of the surface brightness profile differences
between the 2MASS project and our study for 421 elliptical galaxies.  The
differences are primarily found in the outer regions, increasing with
galaxy radius.  The differences are uncorrelated with the luminosity of the
galaxy, size or any other physical characteristic that we can determine.
}
\end{figure}

And the discrepant surface brightness profiles for NGC 3379 is not unique.
The profile differences for all 31 galaxies is shown in Figure
14, presented as a density distribution of $\Delta\mu$ versus
radius.  As can be seen in that Figure, all the comparison galaxies have
varying degrees of surface brightness differences, mostly concentrated in
the outer regions and can reach 1 to 2 mags arcsecs$^{-2}$ in error.

\subsection{Data Reduction Differences}

One obvious conclusion is that some difference exists in the reduction
process that reflects in the final profiles, the data frames themselves are
not in question since we can reproduce 2MASS's aperture luminosities.
There are several procedural differences between the isophotal techniques
used by the 2MASS project and our photometry package (ARCHANGEL).

\begin{figure}[!ht]
\centering
\includegraphics[scale=0.8]{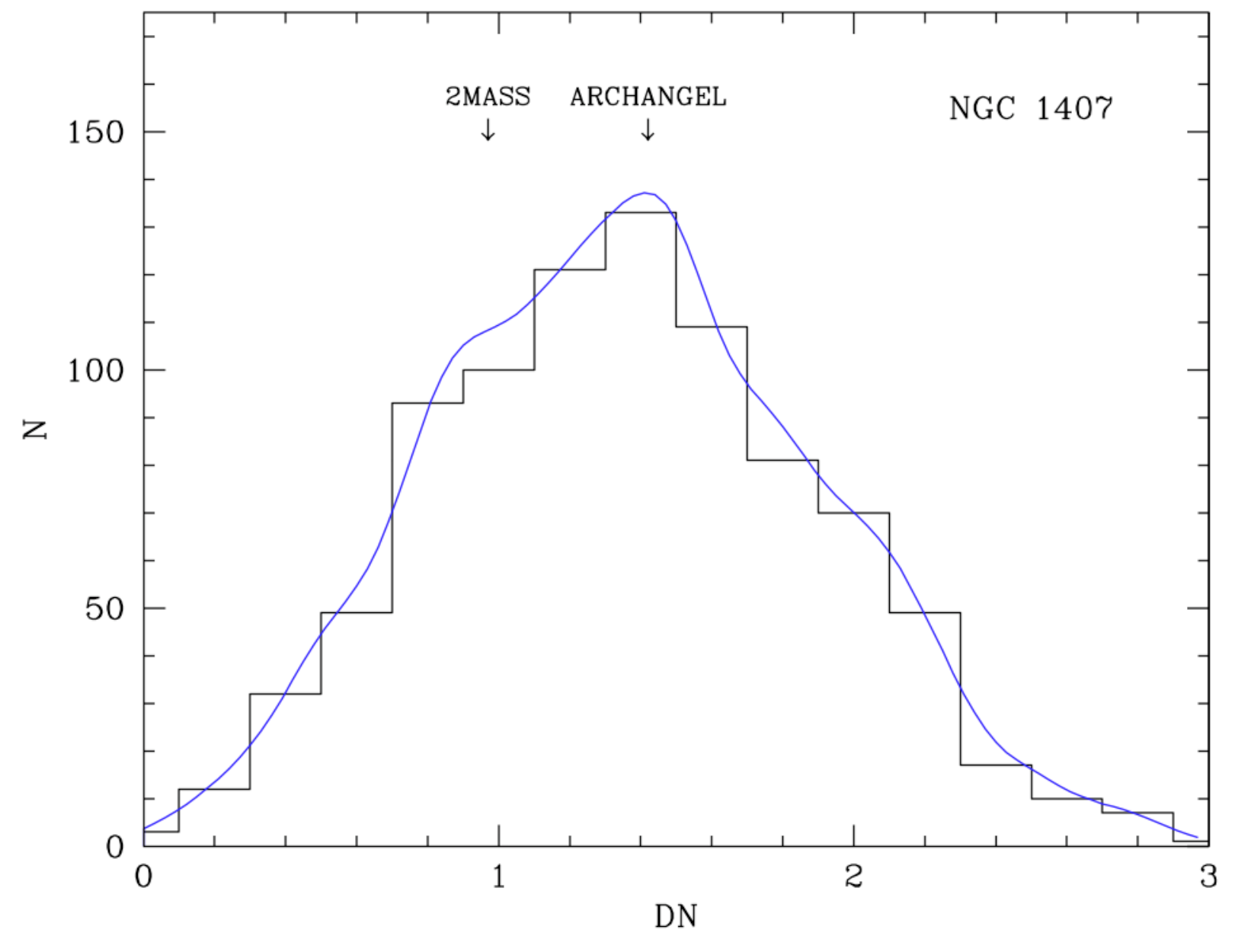}
\caption{\small A histogram of intensity values (in raw data units) for an annulus of
100 arcsecs (width of one pixel) for NGC 1407.  The 2MASS project cites a
value of 0.97 for this annulus, our study finds an intensity of 1.42.  The data
clearly supports our higher value.  This type of test was completed for all
31 galaxies in our surface brightness overlap sample, all produced the same
result.}
\end{figure}

For example, the 2MASS project determines an elliptical shape based on a
first moment analysis of some intermediate, but high S/N region in a
galaxy's envelope.  The calculated eccentricity and position angle are used
for the entire galaxy, determining isophote intensity levels based on
pixels around those ellipses.  Our project, on the other hand, fits each
radii for eccentricity and position angle (as well as x and y center)
allowing these ellipse parameters to vary with radius.

This difference in ellipse shape was noted in Schombert (2007), but these
different ellipses parameters are not sufficient explain the large surface
brightness differences found in the galaxy sample (numerical experiments
with ellipses in 2MASS data displays only a 1 to 2\% difference in
intensities).  There are a few extreme cases (e.g. LSB galaxy, NGC 3109),
but in general ellipticals have fairly constant eccentricities.

In one-to-one comparisons to the raw intensity files provide by the 2MASS
project, one can see there are large differences in the quoted intensity
values per radius between the 2MASS project and our study.  These
differences range from small to up to 60\%, largest at the lowest intensity
values.  An example is outlined in the next section.

\subsection{NGC 1407: A Test Case}

To resolve the differences in the surface brightness profiles, elliptical
NGC 1407 was selected for more detailed inspection.  NGC 1407 is an
excellent test galaxy for its isophotes are nearly circular (axial ratio of
0.93 from 2MASS, 0.95 from our study), it is isolated with no large
companions and its envelope is free of any foreground stars or distortions.

\begin{figure}[!ht]
\centering
\includegraphics[scale=0.8]{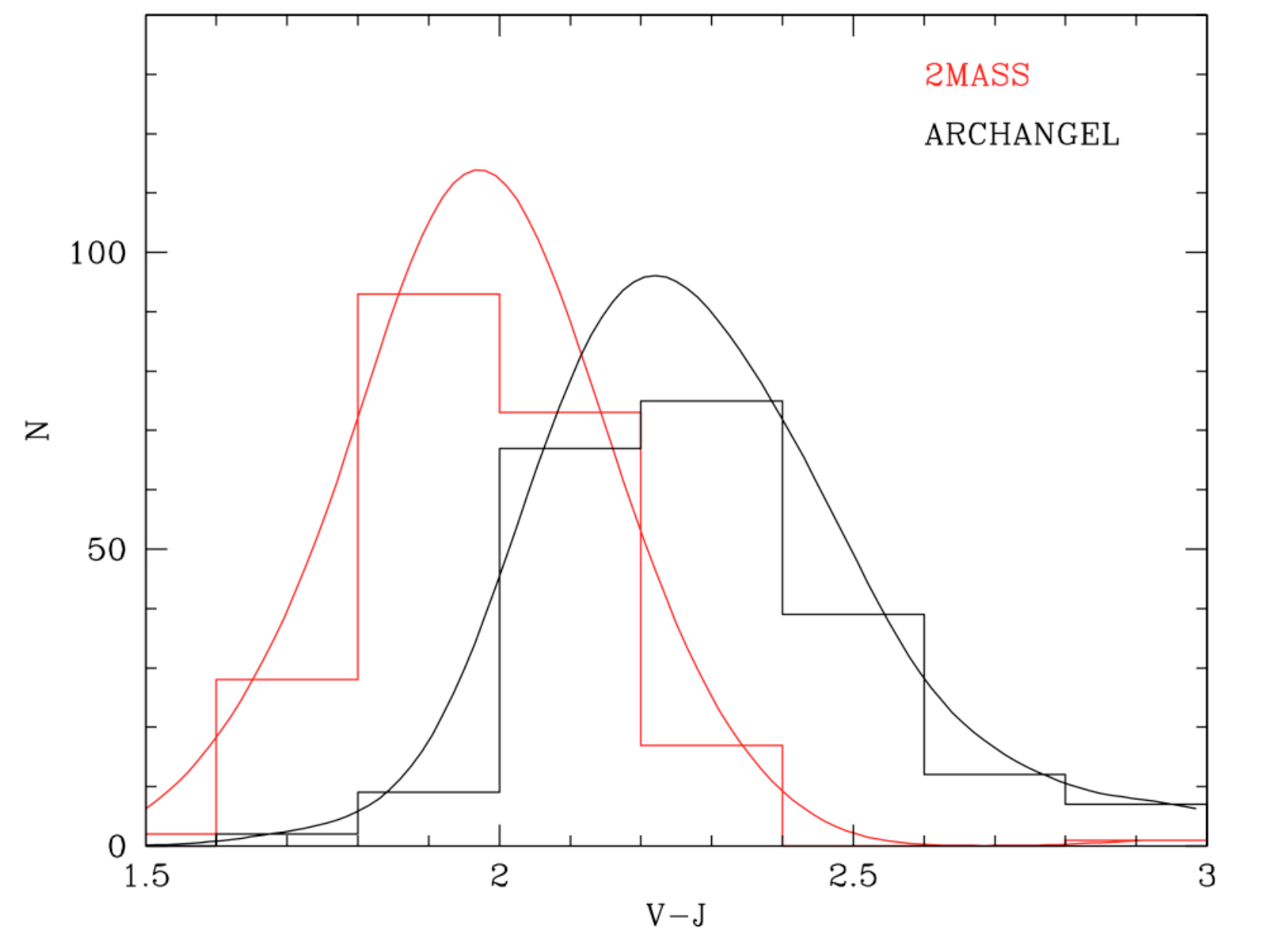}
\caption{\small A histogram of $V-J$ colors using galactic extinction
corrected total magnitudes from 2MASS (extracted from NED) and our study.
Since the 2MASS project underestimates the total magnitudes, this
reflects into bluer $V_J$ compared to our colors.  SED models predict a
$V-J$ color of 2.5 for a solar metallicity stellar population with an age
of 13 Gyrs, in-line with our colors.}
\end{figure}

At 100 arcsecs from the center of NGC 1407, the 2MASS project quotes an
isophotal intensity of 0.97 DN (20.82 $J$ mag arcsecs$^{-2}$).  Our project
finds a value of 1.42 DN (20.41 $J$ mag arcsecs$^{-2}$).  To determine
which value more closely represents the isophote at that radius, we have
plotted a histogram of intensity values for all pixels between 99.5 and
100.5 arcsecs from the galaxy center.  This histogram is shown in Figure
15 (both regular and normalized).

From this Figure, it is obvious that the intensity values deduced by the
2MASS project are not in agreement with the mean value of the pixels in the
image, whereas our calculated intensity value is in good agreement with the
mean and median value.  Since the isophotes of NGC 1407 are nearly a
perfect circle, this is not an effect of the ellipse fitting procedure.
This is also not due to calibration errors, as these are raw data numbers.

Comparison to other isophotes reveals the same difference, always at a
constant value in intensity at all radii and suggests a additive error in
the deduced sky value.  Communication with the 2MASS project (Jarrett 2011)
confirms that the difference to the 2MASS surface brightness profiles is
due to an error in the sky subtraction scheme.  Simply adding a constant
value to the raw intensities (e.g. 0.22 DN for NGC 3379) results in a good
agreement between our current profiles and 2MASS LGA (see Figure
13.

\begin{figure}[!ht]
\centering
\includegraphics[scale=0.8]{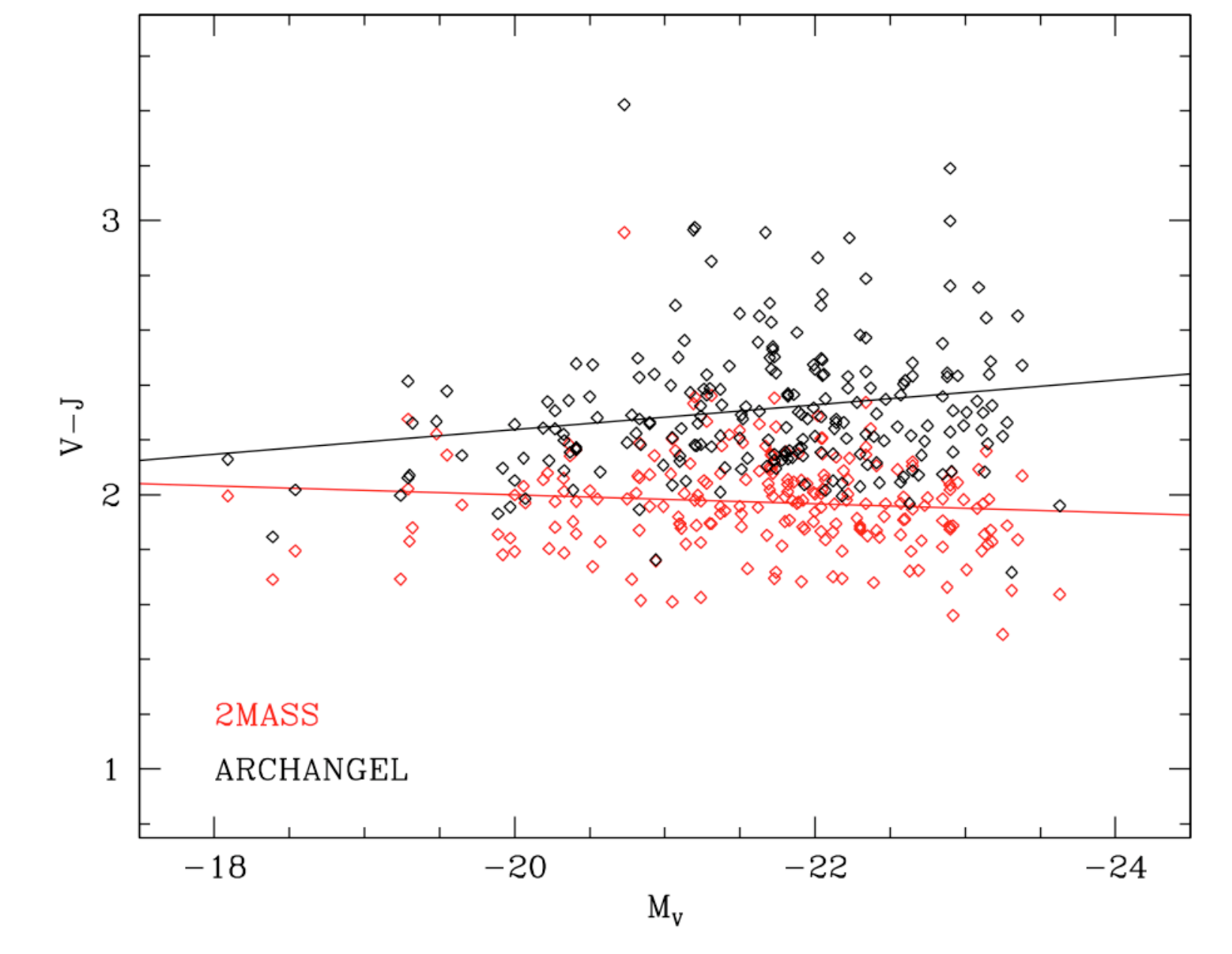}
\caption{\small The $V-J$ color-magnitude diagram for 2MASS colors (red
symbols) versus our study (black symbols).  Linear fits are shown.  The
2MASS project predicts a positive CMR slope, in contradiction with known
negative slopes in the literature.  Our study finds a negative slope
(redder colors with higher galaxy mass, i.e., higher mean metallicity).}
\end{figure}

The effect these underestimated surface brightness values have on 2MASS
photometry is subtle.  Both 2MASS Kron and total magnitudes use the surface
brightness profiles to deduce isophotal levels (Kron) and scalelengths
(total).  For Kron magnitudes, the 20 $K$ mag arcsecs$^{-2}$ level is used
to define an elliptical aperture.  However, since the 2MASS surface brightness
profiles underestimate the intensity values per radius, this, in turn,
leads to smaller estimates of the isophotal size of the aperture and,
therefore, fainter magnitudes.  

Total magnitudes for 2MASS are calculated using an outer aperture set to be
four times the scalelength determined by S\'{e}rsic function fits.
Decreased intensities in the outer regions produce smaller scalelengths, on
average, which produce smaller apertures and fainter total magnitudes.
This is exactly what we observe in Figure 12.

\subsection{Summary}

First, we note that this discrepancy has no impact on projects which use
2MASS aperture colors.  For galaxy colors are calculated using 2MASS total
magnitudes still use the same sized apertures for $J$, $H$ and $K$, and
the colors will remain consistent (although for a smaller portion of the
total galaxy light).  However, comparison between other total magnitudes
(e.g. RC3 magnitudes) and 2MASS total magnitudes will be biased
towards the blue.

\begin{figure}[!ht]
\centering
\includegraphics[scale=0.7]{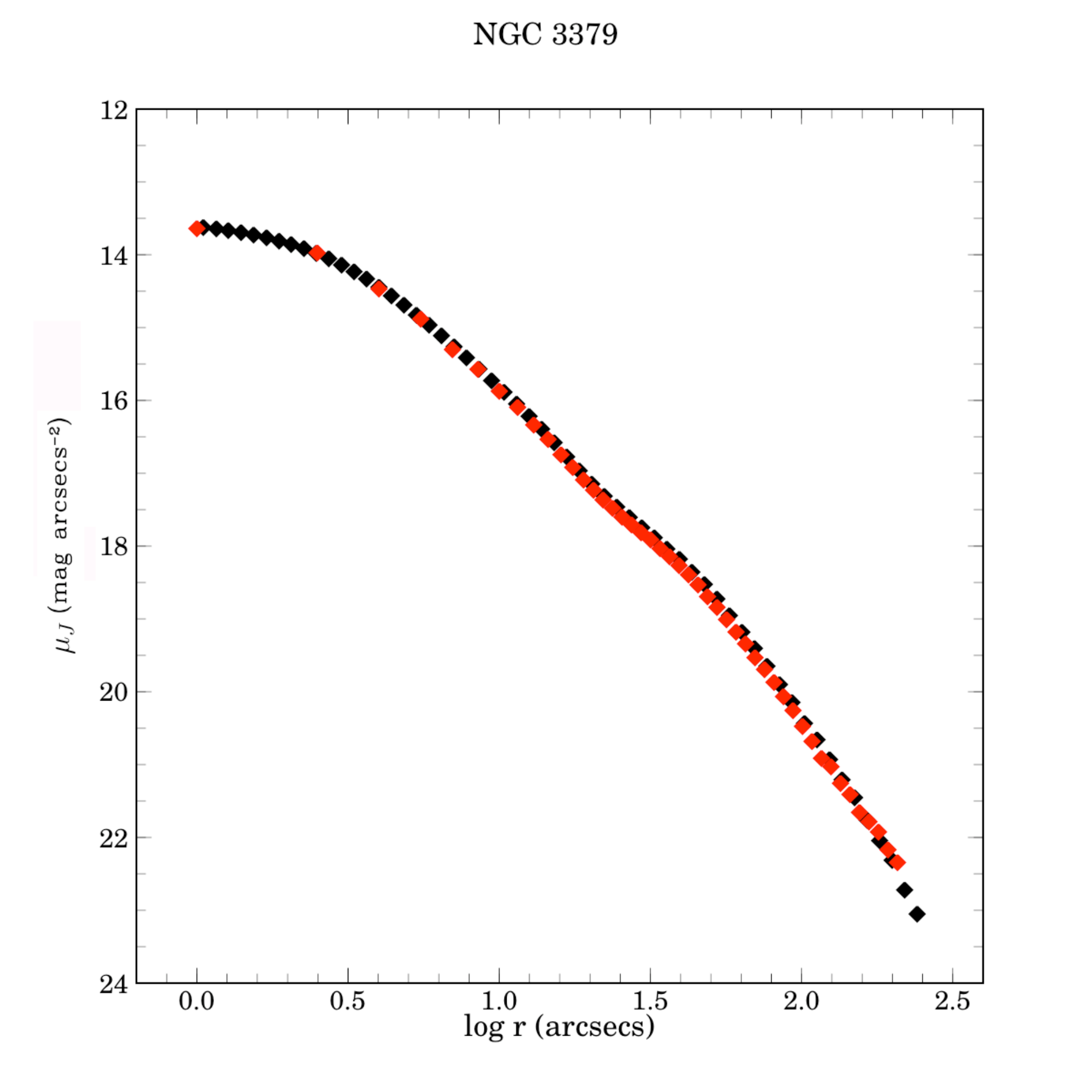}
\caption{\small The effect of adding a small constant value to 2MASS LGA
brightness profiles.  The black data is from this study, the red is the
corrected 2MASS LGA profile with an added value of 0.4\% (0.22 DN).  This
extreme small change brings both profiles back into agreement.}
\end{figure}

An example of this effect is shown in Figure 16, a histogram of
$V-J$ colors for the 421 ellipticals in our sample.  As can be seen, the
2MASS colors are 0.25 mags bluer than colors calculated from our total
magnitudes since the RC3 $V$ magnitude is determined from an asymptotic fit
and, therefore, contains more flux that 2MASS's total magnitude.   There
appears to be no standard correction from 2MASS colors to the correct
colors, this would require information on how deviant the 2MASS surface
brightness profiles (from which the aperture sizes are extracted) are from
reality.

A priority science goal for 2MASS was large baseline color comparison.  An
example of relevance of large wavelength comparisons is the color-magnitude
relation (CMR).  The CMR is a long known correlation between galaxy color
and luminosity.  The best explanation is that galaxies with higher mass
have higher metallicities.  Global metallicity reflects in the mean
temperature of the RGB such that low metallicities produce bluer colors.
The CMR for this data sample is shown in Figure 17.  Again, we
see that the 2MASS colors predict the {\it opposite} expectation from
earlier optical work in that they find roughly bluer colors with higher
luminosity.  Using our total magnitudes (combined with RC3 colors) restores
the correct CMR, redder colors with higher galaxy luminosity.

\section{Asymptotic Total Magnitude Fitting}

Often the scientific goal of a galaxy project is to extract a total
luminosity for the system (and colors for multiple filters).  For small
galaxies, a metric aperture or isophotal magnitude is suitable for
comparison to other samples (certainly the dominate source of error will
not be the aperture size).  However, for galaxies with large angular size
(i.e. many pixels), their very size makes total luminosity determination
problematic.

Naively, one would think that a glut of pixels would make the problem of
determining a galaxies luminosity easier, not more difficult.  However, the
problem here arises with the question of assigning a point where the
galaxy luminosity ends.  Or, even if one estimates or calculates an outer
radius, does the luminosity estimate contain all the galaxy's light.  The
solution proposed by de Vaucouleurs' decades ago is to use a curve of
growth (de Vaucouleurs 1977).  A majority of galaxies follow either an
exponential or r$^{1/4}$ curve of growth such that the total light of a
galaxy can be calculated (Burstein \etal 1987).  However, for modern large
scale CCD imaging, the entire galaxy can easily fit onto a single frame and
there is no need for a curve of growth as all the data exists in the frame.

\begin{figure}[!ht]
\centering
\includegraphics[scale=0.5]{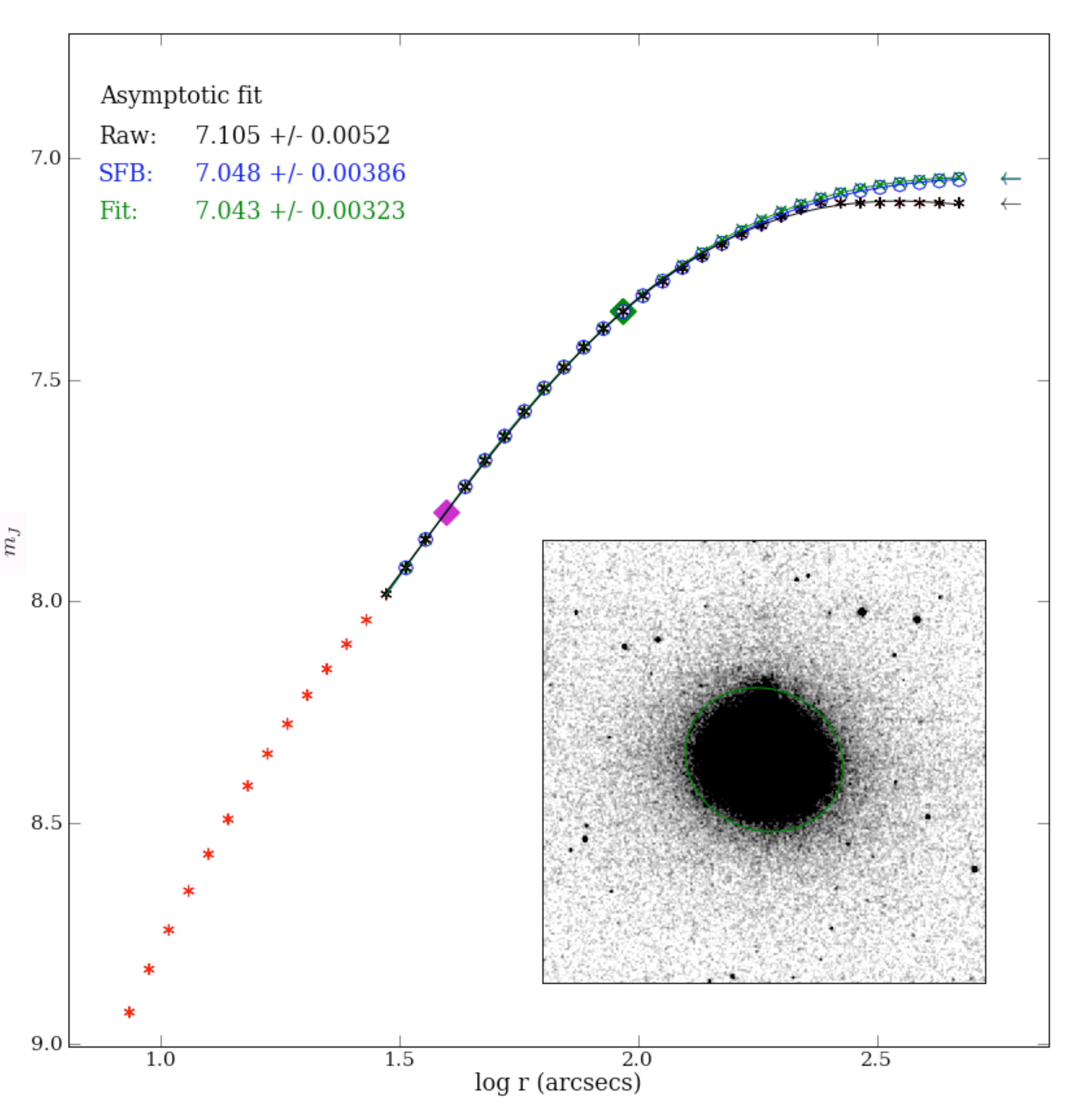}
\caption{\small A plot of the curve of growth for NGC 3379 using elliptical
apertures that follow the isophotes.  The black symbols are the raw
intensities, blue symbols are aperture intensities determined from using
isophote intensities (integrated starting at the point marked by the green diamond)
and green symbols are aperture intensities determined from the fits to the
surface brightness profiles.  Due to the high quality of the fit to the
surface brightness profile, SFB and Fit data are nearly identical.  It is a
choice of the user on which total magnitude to use for analysis.
}
\end{figure}

With adequate S/N, it would seem to be a simple task to place a
large aperture around the galaxy and sum the total amount of light (minus
the sky contribution).  However, in practice, a galaxy's luminosity
distribution decreases as one goes to larger radii, when means the sky
contribution (and, thus, error) increases.  In most cases, larger and
larger apertures simply introduce more sky noise (plus faint stars and
other galaxies).  And, to further complicate matters, the breakover point
in the optical and near-IR, where the galaxy light is stronger than the sky
contribution will not contain a majority of the galaxy's light (see Figure
7).  So the choice of a higher accuracy inner radius will
underestimate the total light.

The procedure selected in this study, after some numerical experimentation,
is to plot the aperture luminosity as a function of radius and attempt to
determine a solution to an asymptotic limit of the galaxy's light.  This
procedure begins by summing the pixel intensities inside the various
ellipses determined by the ellipse fitting routines.  For small radii, a
partial pixel algorithm is used to determine aperture luminosity (using the
surveyors technique to determine each pixel's contribution to the
aperture).  At larger radii, a simple sum of the pixels, and the number
used, is used.  In addition, the intensity of the annulus based on the
ellipse isophote and one based on the fit to the surface photometric
profile are also outputted at each radii.

Note that a correct aperture luminosity calculation requires that
both an ellipse fit and a 1D fit to the resulting surface photometry
has to have been made. The ellipse fit information is required as these ellipses
will define the apertures, and masked pixels are filled with intensities
given by interpolation of the nearest ellipse. A surface photometric fit
allows the aperture routine to use a simple fit to the outer regions as a
quick method to converge the curve of growth.  The end result is three
possible values for aperture luminosity as a function of radius, 1) raw
pixel counts, 2) pixel counts determined by the mean isophote and 3) pixel
counts determined by the mean surface brightness from fits to the galaxy
surface brightness profile.

Once the aperture luminosities are calculated, there are two additional
challenges to this procedure. The first is that an asymptotic fit
is an unstable calculation to make as the smallest errors at large
radii reflect into large errors for the fit. Two possible solutions
are used to solve this dilemma. The first solution is to fit a 2nd
or 3rd order polynomial to the outer radii in a luminosity versus
radius plot. Most importantly for this fit, the error assigned to the
outer data points is the error on the knowledge of the sky, i.e. the
RMS of the mean of the sky boxes. This is the dominant source of error
in large apertures and the use of this error value results in a fast
convergence for the asymptotic fit. The resulting values from the
fit will be the total magnitude and total isophotal size, determined
from the point where the fit has a slope of zero. 

A second solution is to use an obscure technique involving rational
functions. A rational function is the ratio of two polynomial functions of
the form

\[
f(x)=\frac{{a_{n}x^{n}+a_{n-1}x^{n-1}+...+a_{2}x^{2}+a_{1}x+a_{0}}}{{b_{m}x^{m}+b_{m-1}x^{m-1}+...+b_{2}x^{2}+b_{1}x+1}}
\]

\noindent where $n$ and $m$ are the degree of the function. Rational
functions have a wide range in shape and have better interpolating
properties than polynomial functions, particularly suited for fits
to data where an asymptotic behavior is expected. A disadvantage is
that rational functions is their non-linear behavior which, when unconstrained,
will produce vertical asymptotes due to roots in the denominator
polynomial. A small amount of experimentation found that the best rational
function for aperture luminosities is the quadratic/quadratic form, meaning
a degree of two in the numerator and denominator. This is the simplest
rational function and has the advantage that the asymptotic magnitude is
simply $a_{2}/b_{2}$, although is best evaluated at some radii in the halo
of the galaxy under study.

\begin{figure}[!ht]
\centering
\includegraphics[scale=0.8]{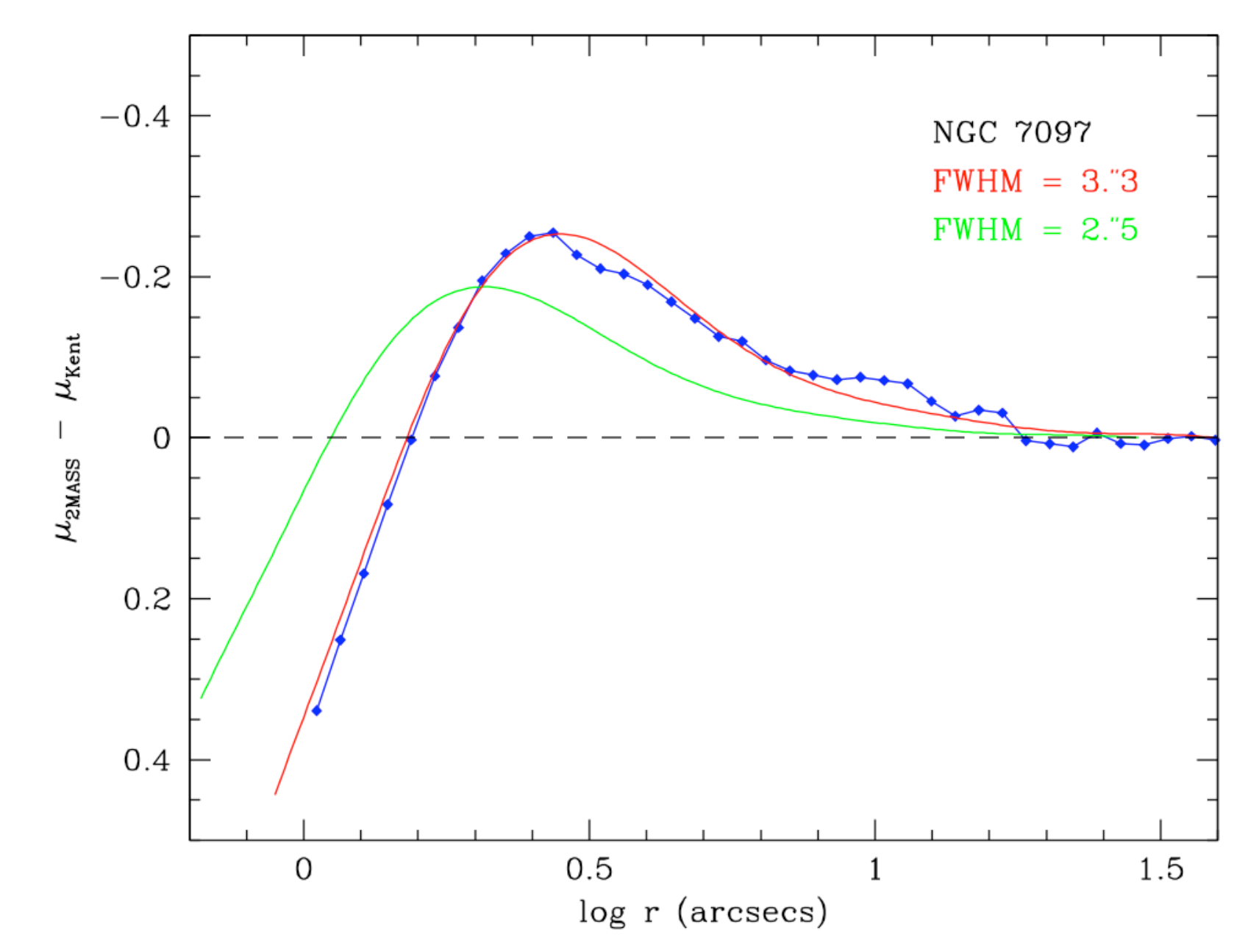}
\caption{\small The PSF for NGC 7097 determined from subtracting sub-arcsec
seeing profile from Kent (2011) and the 2MASS profile (blue symbols).  The
red curve is the best fit Saglia \etal PSF for an $r_e$ of 8.5 arcsecs
resulting in a FWHM of 3.3 arcsecs (the estimated FWHM from the raw frames
was 3.1 arcsecs).  The green curve displays the
sensitivity of PSF fitting, where the same $r_e$ is used, but a FWHM of 2.5
arcsecs (as quoted by the 2MASS project) is assumed.  The PSF can be
recovered, but blind fitting is very sensitive to the input parameters.}
\end{figure}

Usually aperture luminosity values do not converge at the outer edges of a
galaxy. This is the second challenge to aperture photometry, correct
determination of the total luminosity due to the faint galaxy halo
component.  In this instance, the surface photometry profile is critical in
determining the total flux.  Contained in the surface brightness profile of
a galaxy is the relationship between isophotal luminosity and radius, using
all the pixels around the galaxy. The isophotal intensity times the area of
an annulus is often a more accurate number than attempting to determine the
integrated luminosity in an annulus by summing pixels. 

The isophotal information can be used to constrain the curve of growth in
two ways. One, we can use the actual surface brightness intensities and
convert them to a luminosity for each annulus by multiplying the mean
intensity times the area of the annulus. Then, this value can be compared
to the aperture value and flagged where the two begin to radically deviate.
Sometimes, for particularly low surface brightness halos, even the
isophotal intensities will vary at large radii and, thus, a second, more
stable method is to make a linear fit of an exponential, r$^{1/4}$ or
combined function to the outer radii and interpolate/extrapolate that fit
to correct the aperture numbers.

Figure 19 displays the results for all three techniques
for the galaxy NGC 3379. The black symbols are the raw intensities summed
from the image file. The blue symbols are the intensities determined from
the surface photometry. The green symbols are the intensities determined
from the fits to the surface photometric profile.  Since the surface
brightness profile of NGC 3379 is well fit by a S\'{e}rsic function, the
curve of growth using the actual surface brightness data and the fit are
nearly identical.  Typically the raw intensities profile falls below the
surface brightness intensities due to losses from masked pixels.  This is
the case for NGC 3379,
the aperture values fall below the surface brightness intensities to
produce a fainter total magnitude.

\section{2MASS PSF Fitting}

While the 2MASS PSF is well known, removing the effects of PSF distortion
involves a assumption to the underlying galaxy profile.  For a majority of
galaxies (ellipticals and spirals with significant bulges), an r$^{1/4}$
profile is a good approximation for PSF correction.  However, the effective
radius must be included in the PSF fit and can only be estimated by
r$^{1/4}$ fits to the region outside the PSF.  Therefore, PSF correction,
using an FWHM measured from nearby stars and an $r_e$ value measured from
the middle portions of a galaxy profile, simply converts the inner profile
into an r$^{1/4}$ shape that is an extension to the middle regions (where
the value of $r_e$ is determined).

Whether this technique is appropriate can be tested by comparing 2MASS
profiles with surface brightness profiles extracted from pointed
observations with higher spatial resolution.  During a project to study the
luminosity function of galaxies in the near-IR, Kent (2011) imaged several
of our the galaxies from our elliptical sample in sub-arcsec seeing with 0.2
arcsec pixels.  The profile for NGC 7097 was shown in Figure 
6.

As a test of the observed PSF, we can subtract the Kent profile from the
2MASS profile for NGC 7097, shown in Figure 20 as the blue
symbols.  Note that the resulting data curve matches the gaussian shape
from Saglia \etal.  To find a best match, one only needs the effective
radius and the seeing FWHM.  For this exercise, we have fit the profiles
outside the seeing region and determined that both profiles have $r_e$
values of 8.6 arcsecs.

Two curves for FWHM of 2.5 and 3.3 arcsecs are shown.  The value of 2.5 is
extracted from the FITS header for the central frames header (taken from
the output of the 2MASS project's SEEMAN routine).  The fit for the 2.5
arcsecs seeing does not fit the data.  A good fit is found for a FWHM of
3.3 arcsecs (which matches our estimate of the 2MASS seeing from examining
stars in the same data frame), which supports the hypothesis that 2MASS PSF
is a gaussian, although the seeing FWHM estimate from 2MASS appears to
underestimate the true seeing.  The true value can be extracted from stars
in the same image frame, and supports our conservative policy of only using
data outside 5 arcsecs for profile fitting.

\section{ARCHANGEL XML Parameter List}

The information for the reduction pipeline, and the resulting structural
and photometric parameters, are contained in each galaxy's XML file.  The
following is a short description of each of those values.

\begin{deluxetable}{ll}
\tablecolumns{2}
\small
\tablewidth{6.0in}
\tablecaption{XML Variable Names}
\tablehead{
\colhead{Variable Name} & \colhead{Description} \\
}
\startdata

\sidehead{Data Informaton:}

{\tt origin} & source of dataset (e.g., 2MASS, SDSS)  \\
{\tt morph\_type} &  morphological type of galaxy from RSA/UGC  \\

\sidehead{Structural Parameters:}

{\tt re\_dev} & effective radius (r$^{1/4}$ fit)  \\
{\tt se\_dev} & effective surface brightness (r$^{1/4}$ fit) \\
{\tt lower\_fit\_dev} & lower fitting radius (r$^{1/4}$ fit) \\
{\tt upper\_fit\_dev} & upper fitting radius (r$^{1/4}$ fit) \\
{\tt chisq\_dev} & $\chi^2$ (r$^{1/4}$ fit) \\
{\tt re\_bulge} & effective radius (bulge+disk fit) \\
{\tt se\_bulge} & effective surface brightness (bulge+disk fit) \\
{\tt chisq\_bulge} & $\chi^2$ (bulge fit) \\
{\tt mu\_o} & central surface brightness (bulge+disk fit) \\
{\tt alpha} & disk scalelength (bulge+disk fit) \\
{\tt lower\_fit\_disk} & lower fitting radius (bulge+disk fit) \\
{\tt upper\_fit\_disk} & upper fitting radius (bulge+disk fit) \\
{\tt chisq\_disk} & $\chi^2$ (disk fit)  \\
{\tt mu\_c} & raw central surface brightness \\
{\tt bdratio} & bulge to disk ratio (luminosity units) \\
{\tt re\_sersic} & effective radius (S\'{e}rsic fit) \\
{\tt se\_sersic} & effective surface brightness (S\'{e}rsic fit) \\
{\tt n\_sersic} & power-law index (S\'{e}rsic fit) \\
{\tt lower\_fit\_sersic} & lower fitting radius (S\'{e}rsic fit) \\
{\tt upper\_fit\_sersic} & upper fitting radius (S\'{e}rsic fit) \\
{\tt chisq\_sersic} & $\chi^2$ (S\'{e}rsic fit)  \\
{\tt template\_mag} & best fit template magnitude \\
{\tt template\_sig} & $\sigma$ on template fit \\

\enddata
\end{deluxetable}

\begin{deluxetable}{ll}
\tablecolumns{2}
\small
\tablewidth{6.0in}
\tablenum{1-Continued}
\tablecaption{XML Variable Names}
\tablehead{
\colhead{Variable Name} & \colhead{Description} \\
}
\startdata

\sidehead{Photometric Parameters:}

{\tt tot\_mag\_raw} & total magnitude using raw intensities \\
{\tt tot\_mag\_raw\_err} & error in total magnitude using raw intensities \\
{\tt tot\_rad\_raw} & total radius using raw intensities \\
{\tt tot\_mag\_raw\_last} & last raw intensity used in total magnitude \\
{\tt tot\_mag\_fit} & total magnitude using fit intensities \\
{\tt tot\_mag\_fit\_err} & error in total magnitude using fit intensities \\
{\tt tot\_rad\_fit} & total radius using fit intensities \\
{\tt tot\_mag\_sfb} & total magnitude using isophotal intensities \\
{\tt tot\_mag\_sfb\_err} & error in total magnitude using isophotal intensities \\
{\tt tot\_rad\_sfb} & total radius using isophotal intensities \\
{\tt tot\_mag\_half\_lum} & half total luminosity \\
{\tt tot\_mag\_half\_rad} & radius at half luminosity point \\
{\tt tot\_mag\_iter\_pt} & radius for iteration of isophotal and fit intensities \\
{\tt tot\_mag\_quality} & note on if total magnitude converged \\
{\tt tot\_mag\_sky} & sky value used for total mags if different from sky\_box value \\

\enddata
\end{deluxetable}

\begin{deluxetable}{ll}
\tablecolumns{2}
\small
\tablewidth{6.0in}
\tablenum{1-Continued}
\tablecaption{XML Variable Names}
\tablehead{
\colhead{Variable Name} & \colhead{Description} \\
}
\startdata

\sidehead{Data Arrays:}

{\tt prf} & ellipse fitting results \\

  ~~~ {\tt INTENS} & intensity (in DN units) \\
  ~~~ {\tt INT\_ERR} & error in intensity \\
  ~~~ {\tt GRAD} & slope of intensity gradient \\
  ~~~ {\tt RAD} & semi-major axis in pixel units \\
  ~~~ {\tt RMSRES} & RMS residuals around ellipse \\
  ~~~ {\tt FOURSL} & some measure of changing slope \\
  ~~~ {\tt ITER} & number of iterations \\
  ~~~ {\tt NUM} & number of pixels \\
  ~~~ {\tt RESID\_1} & residuals on 1st component \\
  ~~~ {\tt RESID\_2} & residuals on 2nd component \\
  ~~~ {\tt RESID\_3} & residuals on 3rd component \\
  ~~~ {\tt RESID\_4} & residuals on 4th component \\
  ~~~ {\tt ECC} & ellipticity \\
  ~~~ {\tt POSANG} & position angle \\
  ~~~ {\tt X0} & x center \\
  ~~~ {\tt Y0} & y center \\
  ~~~ {\tt FOUR\_2} & Fourier quotient \\
  ~~~ {\tt THIRD\_2} & Fourier quotient \\

\enddata
\end{deluxetable}

\begin{deluxetable}{ll}
\tablecolumns{2}
\small
\tablewidth{6.0in}
\tablenum{1-Continued}
\tablecaption{XML Variable Names}
\tablehead{
\colhead{Variable Name} & \colhead{Description} \\
}
\startdata

\sidehead{Data Arrays:}

{\tt colors} & array of $J$,$H$,$K$ colors \\
  ~~~ {\tt radius} & semi-major axis radius in pixel units \\
  ~~~ {\tt J-K} & $J-K$ color \\
  ~~~ {\tt err\_(J-K)} & error in $J-K$ color \\
  ~~~ {\tt H-K} & $H-K$ color \\
  ~~~ {\tt err\_(H-K)} & error in $H-K$ color \\

{\tt ept} & aperture photometry \\

  ~~~ {\tt radius} & semi-major axis radius in pixel units  \\
  ~~~ {\tt mag} & -2.5log(DN) \\
  ~~~ {\tt area} & number of pixels \\
  ~~~ {\tt xsfb} & correction from isophotal intensities \\
  ~~~ {\tt expm} & correction from fit intensities \\
  ~~~ {\tt kill} & deletion flag \\

{\tt sfb} & surface brightness data \\

  ~~~ {\tt radius} & semi-major axis radius in pixel units \\
  ~~~ {\tt mu} & surface brightness in mags arcsecs$^{-2}$ \\
  ~~~ {\tt kill} & deletion flag \\
  ~~~ {\tt error} & error in surface brightness \\

{\tt sky\_boxes} & position of sky boxes \\

  ~~~ {\tt x} & pixel x center \\
  ~~~ {\tt y} & pixel y center \\
  ~~~ {\tt box\_size} & box size \\

\sidehead{Calibration Information:}

{\tt zeropoint} & photometric zeropoint \\
{\tt scale} & plate scale (arcsecs per pixel) \\
{\tt sky} & sky value determined by sky boxes \\
{\tt skysig} & $\sigma$ on sky value \\
{\tt luminosity\_distance\_moduli} & distance moduli (from NED, either CMB or non-redshift value) \\
{\tt gal\_extinc\_J} & galactic extinction in $J$  \\
{\tt gal\_extinc\_H} & galactic extinction in $H$ \\
{\tt gal\_extinc\_K} & galactic extinction in $K$ \\
{\tt magnitude} & \\

\enddata
\end{deluxetable}

\pagebreak

\end{document}